\documentclass[11pt]{article}

\usepackage[usenames,dvipsnames]{color}            
\newcommand{\wf}[1]{}
\newcommand{\wk}[1]{}

\usepackage{graphicx,epsfig,amssymb,amsbsy,amsmath,amsthm,cleveref,enumitem}
\newcommand{\bb}[1]{\left({#1}\right)}					
\newcommand{\sq}[1]{\left[#1\right]}						
\newcommand{\cc}[1]{\left\{#1\right\}}					
\newcommand{\op}[1]{\mathcal{#1}}
\newcommand{\ord}[1]{{\sf O}\mbox{\small$\bb{#1}$}}					
		
\newcommand{\sfrac}[2]{\mbox{$\frac{#1}{#2}$}}	
\newcommand{\hf}{\mbox{$\sfrac12$}}

\renewcommand{\v}[1]{{\bf #1}}							

\newcommand{\uuline}[1]{\underline{\underline{#1}}}

\newcommand{\im}{{\operatorfont i}}									
\newcommand{\id}{1}				

\newcommand{\omegab}{{\mbox{\boldmath$\omega$}}}
\newcommand{\eps}{\varepsilon}

\newcommand{\mirror}[1]{\overline{#1}}
\newcommand{\conj}[1]{{{#1}^*}}
\newcommand{\adj}[1]{{{#1}^\dagger}}

\newcommand{\reals}{\mathbb{R}}						
\newcommand{\imags}{\mathbb{I}}						
\newcommand{\coms}{\mathbb{C}}						
\newcommand{\rasps}{\mathbb{J}}
\newcommand{\rep}[1]{\reals\sq{#1}}					
\newcommand{\imp}[1]{\imags\sq{#1}}					

\renewcommand{\exp}[1]{\operatorname{e}^{#1}}
\newcommand{\e}[1]{{\mbox{\boldmath$e$}}_{#1}}						
\newcommand{\ec}[1]{{\mbox{\boldmath$e$}}^{#1}}						
\newcommand{\tayl}{Y}								
\newcommand{\cris}{\Gamma}							
\newcommand{\con}{\Gamma}
\newcommand{\scon}{{ P}}							
\newcommand{\tens}{\uuline}

\renewcommand{\tayl}[2]{{\sf tayl}}

\newcommand{\m}{M}
\newcommand{\n}{N}

\newcommand{\z}{z }

\newtheorem{theorem}{Theorem}
\newtheorem{lemma}[theorem]{Lemma}

\newtheorem{example}{Example}

\crefname{equation}{}{}
\Crefname{equation}{}{}

\def\eps{\varepsilon}

\begin{document}

\title{Complex geometry and fundamental physical law}
\author{Mike R. Jeffrey\footnote{University of Bristol, UK}}
\date{\today}

\maketitle

\begin{abstract}\noindent
We present here a product between vectors and scalars that {\it mixes} them within their own space, using imaginaries to describe geometric products between vectors as complex vectors, rather than introducing higher order/dimensional vector objects. This is done by means of a {\it mixture tensor} that lends itself naturally to tensor calculus. We use this to develop a notion of analyticity in higher dimensions based on the idea that a function can be made differentiable --- in a certain strong sense --- by permitting curvature of the underlying space, and we call this {\it analytic curvature}. 

To explore these ideas we use them to derive a few fundamental laws of physics which, while considered somewhat lightly, have nevertheless compelling features. The mixture, for instance, produces rich symmetries without adding dimensions beyond the familiar space-time, and its derivative produces familiar quantum field relations in which the field potentials are just derivatives of the coordinate basis. 
\end{abstract}

\newpage

\tableofcontents

\newpage


\section{From mixtures to fundamental laws}\label{sec:intro}

The aim of this paper is to introduce a product that {\it mixes} vectors within their own space, and to investigate the differentiability of functions within the algebra so created. To illustrate the formalism we will use it to obtain some familiar differential relations of fundamental physical fields, as little more than a mathematical game, but revealing rich symmetries and perturbations of potential interest in physical law. 

We start by defining the product between a general pair of vector bases $\e\beta$ and $\e\gamma$ by means of a {\it mixture} tensor $\eta$ as a vector $\e\alpha=\e\beta\e\gamma=\eta(\e\beta,\e\gamma)$. This can be considered an alternative to geometric or exterior algebras, where instead of products between vectors creating higher `order' quantities such as bivectors associated with areas or rotations, trivectors associated with volumes, and so on, we more simply define products of vectors as producing new vectors of the same order, a `functional' rather than `geometric' algebra, perhaps. Rotations {\it along} a vector and rotations {\it about} a vector direction are distinguished through the important role played by the imaginary $\im=${\small$\sqrt{-1}$}. 

In fact the mixture is a small augmentation of standard concepts, as it unifies symmetric products related to the metric, with antisymmetric products related to `structure coefficients' or Levi-Civita symbols of non-commutative bases; indeed such a combined product may well have been developed elsewhere that the author is not aware of. The real impact of the mixture, however, is on calculus. 

We then ask what happens if we try to express the differential of a function as
\begin{align}\label{df1}
df= dz\sfrac{d\;}{dz}f+\ord{dz^2}\;,\qquad\qquad
\end{align}
such that the function $f$, variable $z$, and derivative $\sfrac{d\;}{dz}f$, are all of the same type (or belong to the same `space'). The key motivation for seeking such a form is so that the integral $\int dz\;g$, of some function $g$ with an antiderivative $f$, can be shown to satisfy
\begin{align}\label{fund}
\int_a^b\!\!dz\;g=\int_a^b\!\!dz\;\sfrac{d\;}{dz}f=\int_{f(a)}^{f(b)}\!\!\!df=f(b)-f(a)\;
\end{align}
and thus be path independent. 
Such a form of integrability would make many powerful complex integral methods, from residues to steepest descents and stationary phases, applicable to vector functions integrated along curves through vector spaces in more general ways than currently possible. This would suggest implications to variational concepts from Feynman's path integrals to wave asymptotics or optimization problems. The original intent of this study was to explore the possiblity of extending phase integral methods in general to non-scalar variables. Here we set out just some basics of the algebra and calculus that arise, using speculatize applications to fundamental physics to see the kinds of relations that result. 

In one dimension the relations \cref{df1}-\cref{fund} are fundamental to differential and integral calculus. 
In two dimensons they are only possible under certain restrictions, most obviously when $z$ is complex and $f$ satisfies the Cauchy-Riemann conditions, as thse permit \cref{df1} to be written without complex conjugate terms. In higher dimensions the form \cref{df1} becomes too restrictive to hold for all but trivial (i.e. constant) functions, {\it unless} we let $\sfrac{d\;}{dz}f$ consist not only of the obvious derivative operator acting on $f$, namely $\sfrac{\partial\;}{\partial z}f$, but also an error term we call $\con(f)$, such that \cref{df1} can be written as
\begin{align}\label{df2}
df=dz\big(\sfrac{\partial\;}{\partial z}f+\con(f)\big)+\ord{dz^2}\;.
\end{align}
We will associate the error term $\con(f)$ with the derivative of the basis in which $f$ and $z$ are expressed. That is, to obtain \cref{df1} in its augmented form \cref{df2}, we will permit variation of the coordinate basis, and in doing so make the requirement of differentiability a {\it source term} for curvature of the underlying space. We refer to this as {\it analytic curvature}. 


Thus the conditions of analytic curvature provide us with a covariant extension of the Cauchy-Riemann conditions to higher dimensions. Investigating some typical expressions leads us to differential equations that look tantilizingly like a variety of fundamental physical laws.

We will show in \cref{sec:dirac} that the covariant derivative of a function $f$ can be written as $\sfrac{df}{dz}=(df)^\gamma\e\gamma$ where
\begin{align}
(df)^\gamma=
\eta^{\gamma\beta}_\alpha f^\alpha_{;\beta}
&=(\eta^{\gamma\beta}_\alpha \partial_{\beta}+ H^\gamma_\alpha)f^\alpha\;,\qquad\qquad\;
\end{align}
from which we will see that analytic curvature provides the Dirac equations \cite{dirac28,diracqm}, with the mixture $\eta$ and curvature term $H=\eta\con$ assuming the role of the Dirac matrices. 
If $\con$ is symmetric this derivative can be written
\begin{align}
\eta^{\gamma\beta}_\alpha f^\alpha_{;\beta}&=\eta^{\gamma\beta}_\lambda(1\partial_{\beta}+\op G_{\beta})^\lambda_\alpha f^\alpha\;,
\end{align}
where $\op G_\mu$ is a square matrix with components $(\op G_\mu)^\alpha_\beta=\con^\alpha_{\mu\beta}$, and $1$ is the identity matrix. This form is consistent with the derivative of the standard model of particle physics if $\op G_\mu=\epsilon\op H_\mu$ where $\op H_\mu$ are field potentials and $\epsilon$ a coupling parameter. The curvature tensor is then
\begin{align}
R^{\alpha}_{\beta\nu\mu}&=2\epsilon(\op H^{}_{[\mu,\nu]}+\epsilon\op H^{}_{[\nu}\op H^{}_{\mu]})^\alpha_\beta
\;:=\;2\epsilon(F_{\mu\nu})^\alpha_\beta\;,
\end{align}
giving the electromagnetic field tensor $F^{\mu\nu}$ of the Yang-Mills theory (see e.g. \cite{cg07,yangmills}). The symmetries and the richness of gauge invariance of physics mixture then arise not from the basis of a space directly, but from the mixture and associated expressions of covariance, requring no dimensions beyond the familar $3+1$ of space-time. We derive these expressions in \cref{sec:standard}. 

In this, the fundamental field potentials are curvature terms required by analyticity. A special case is given if the electromagnetic potential $h_\alpha=\cc{\phi,\v A}$ is just the divergence of the coordinate basis, 
\begin{align}
h_\alpha\ec\alpha=\ec\alpha_{,\alpha} \;,
\end{align}
then the potentials are just the trace of the connection, $h_\alpha=\con^\beta_{\alpha\beta}$, 
and the curvature tensor is related to the electromagnetic field (Faraday) tensor $F_{\mu\nu}$ as 
\begin{align}
R^\gamma_{\alpha\mu\nu}=\id^\gamma_\alpha h_{[\mu,\nu]}=\id^\gamma_\alpha F_{\nu\mu}\;,
\end{align}
a result reminiscent of Pauli's association of the Riemann tensor $R$ with the Faraday tensor $F$. This result itself does not require either the mixture or analytic curvature, but becomes more interesting when we look in more generality in \cref{sec:physics}. 

Imaginary quantities play an inescapable role in the algebra we develop, and will lead us more speculatively to suggest a novel perturbation of flat space that yields a curvature `error' term
\begin{align}
\con^\omega_{\gamma0}\;\;\propto\;\; J^\omega_\gamma\;+\;\hf\im F^{\omega\lambda}\id_{\lambda\gamma}\;,
\end{align}
consisting of a gravitational contribution $J$ and electromagnetic contribution $F$, where
\begin{align}
J^\omega_\gamma=\mbox{\footnotesize$\bb{\begin{array}{cccc}0&G_1&G_2&G_3\\G_1&0&0&0\\G_2&0&0&0\\G_3&0&0&0\end{array}}  $}\;,\quad
F^{\omega\lambda}\id_{\lambda\gamma}=\mbox{\footnotesize$\bb{\begin{array}{cccc}0&E_1&E_1&E_3\\-E_1&0&-B_3&B_2\\-E_2&B_3&0&-B_1\\-E_3&-B_2&B_1&0\end{array}}  $}\;,
\end{align}
with $G=\nabla\psi$, $E=\nabla\phi+\partial_t A$, $B=\nabla\times A$, in terms of the gravitational potential $\psi$ and electromagnetic 4-potential $(\phi,A)$. From this we derive least-variation curves corresponding to geodesics, along which the equations of motion have gravitational component
\begin{align}
cmv^i_{,0}\approx-mG_i +\dots\;,\qquad\qquad\qquad
\end{align}
and electromagnetic component 
\begin{align}
cmv^i_{,0}\approx {e} (E_i + v^j \eta^{ik}_j B_k)+\dots\;,\qquad
\end{align}
for a test particle with mass $m$, charge $e$, and velocity $v$, consistent with Newton's second law in a gravitational field and with the Lorentz force; see \cref{sec:weak}. 
Using the mixture product, the electromagnetic energy flux (Poynting vector) is just the complex magnitude-squared, $|\v E+\im\v B|^2=(\v E+\im\v B)(\v E-\im\v B)$. 
Maxwell's electromagnetic equations, moreover, arise rather easily from the analytic curvature of the electromagnetic fields and potentials.

Of course one may find various elegant expressions of such laws in terms of geometric algebras or other formalisms. The compelling feature of these investigations is how easily the familiar forms of these laws arise in an algebra and calculus based around the {mixture} $\eta$. Besides its original aim of paving the way to new integral methods in higher dimensions, they show how the various possible symmetries of the mixture $\eta$ of bases (rather than the bases themselves) take centre stage in determining the forms of differential laws.

In 
\cref{sec:mixture} we introduce the {\it mixture} as an alternative to geometric algebra, and set out its effects on covariant calculus in 
\cref{sec:covariance}. In \cref{sec:curvature} we introduce the notion of analytic curvature. 
We use these notions to derive some fundamental physical laws in 
\cref{sec:physics}. 
Closing remarks are given in \cref{sec:close}, with some further details in the Appendix, giving a derivation of the natural geometry following an adaptation of Hamilton's quaternions in \cref{sec:Anatgeo}, 
and a few identities concerning the mixture in \cref{sec:Amixture}.

As these ideas are speculative and adrift of any currently conventional directions of research, I eschew many important technicalities of modern differential geometry and proceed somewhat informally. Much room is left for rigour and for connecting to concepts and nomenclatures used across more conventional fields of mathematics and physics. Nevertheless I set the ideas out at enough length, I hope, to suggest just some of the directions they might be developed in. 

\section{Algebra by mixture}\label{sec:mixture}

The following theory takes as a central concept the {\it mixture} of two basis elements to define their algebraic product. 
The idea is essentially that by mutiplying two basis elements $\e\alpha$ and $\e\beta$ in a given space $\op S$, we obtain another element, $\e\gamma\in\op S$, up to multiplication by real or imaginary factors. Rather than employing scalar, vector, and exterior products, we describe such a multiplication via a {\it mixture tensor}, as follows.

\subsection{The mixture of bases}\label{sec:mix}


Given a system of $n$ orthogonal basis vectors $\cc{\e\alpha}_{\alpha=0,1,2,...,n}\;$, some dual basis $\cc{\ec\alpha}_{\alpha=0,1,2,...,n}\;$, and a {mixture tensor} $\eta$ whose components are real or complex scalars, let the products of bases be given by 
\begin{equation}\label{eta}
\e\alpha \e\beta=\eta^\gamma_{\alpha\beta}\e\gamma\qquad\mbox{and}\qquad
\ec\alpha \ec\beta=\eta_\gamma^{\alpha\beta}\ec\gamma\;.\qquad
\end{equation}
We also allow products between vectors and duals, via
\begin{equation}\label{eta2}
\e\beta\ec\alpha=\eta_{\gamma\beta}^\alpha\ec\gamma=\eta_\beta^{\alpha\gamma}\e\gamma\qquad\mbox{and}\qquad
\ec\alpha\e\beta=\eta_{\beta\gamma}^\alpha\ec\gamma=\eta_\beta^{\gamma\alpha}\e\gamma\;.
\end{equation}
We use the convention of summing over repeated upper-lower index pairs (hence summing over $\gamma=0,1,...,n$ in these expressions). 



%

Using the mixture we can decompose the product of two symbols $z=z^\alpha\e\alpha$ and $w=w^\beta\e\beta$ as
\begin{equation}
zw=(\z^\alpha\e\alpha)(w^\beta\e\beta)=\z^\alpha w^\beta \eta^\gamma_{\alpha\beta}\e\gamma:=u^\gamma\e\gamma=u\;\;,
\end{equation}
allowing us to extract the components $u^\gamma=\z^\alpha w^\beta \eta^\gamma_{\alpha\beta}$. This provides closure under multiplication.

If $\eta_{\alpha\beta}^\gamma=\eta_{\beta\alpha}^\gamma$ then the algebra is commutative, and if $\e\alpha(\e\beta\e\gamma)=(\e\alpha\e\beta)\e\gamma$ (or in terms of the mixture $\eta_{\beta\gamma}^\lambda\eta_{\alpha\lambda}^\omega=\eta_{\alpha\beta}^\lambda\eta_{\lambda\gamma}^\omega$) then the algebra is associative. We will mainly consider associative but non-commutative algebras here.  

\bigskip



For convenience we denote symmetrization over indices using round brackets $\bb{..}$, anti-symmetrization using square brackets $\sq{..}$, and cyclic permutation of indices using curly brackets $\cc{..}$, 
\begin{align}
x_{(\mu}y_{\nu)}&=\hf(x_{\mu}y_{\nu}+x_{\nu}y_{\mu})\;,\nonumber\\
x_{[\mu}y_{\nu]}&=\hf(x_{\mu}y_{\nu}-x_{\nu}y_{\mu})\;,\\
x_{\{\mu}y_{\nu}z_{\omega\}}&=x_{\mu}y_{\nu}z_{\omega}+x_{\nu}y_{\omega}z_{\mu}+x_{\omega}y_{\mu}z_{\nu}\;.\nonumber
\end{align}

\subsection{Three operators}

There are three useful operations that help distinguish the scalar/vector and basis/dual parts of a variable:
\begin{itemize}
\item The {\bf mirror} acts on non-commuting bases as
\begin{align}\label{def:mirror}
\begin{array}{ll}
\mirror{\e\alpha}=\e\alpha\qquad&\mbox{if }\e\alpha\mbox{ commutes with all other }\e\beta\;,\\
\mirror{\e\alpha}=-\e\alpha\qquad&\mbox{if }\e\alpha\mbox{ anti-commutes with any other }\e\beta\;,
\end{array}
\end{align}
with no effect on scalars. 

\item The {\bf conjugate} 
commutes bases as
\begin{subequations}\label{def:conj}
\begin{align}\label{def:conje}
\conj{(\e\alpha\e\beta)}=\e\beta\e\alpha\;,
\end{align}
and inverts imaginary scalars as
\begin{align}\label{def:conji}
\;\quad\conj{\im}=-\im\;.
\end{align}
\end{subequations}

\item The {\bf adjoint} is the combination of these, $\adj z=\conj{\mirror z}$, so
\begin{align}\label{def:adjoint}
\adj{(z^\alpha\e\alpha)}=\conj{z^\alpha}\mirror{\e\alpha}\;.
\end{align}

\end{itemize}

Thus the mirror and the conjugate are both anti-commutative, and this makes the adjoint commutative, 
\begin{equation}
\mirror{zw}=\mirror w\;\mirror z\;,\qquad\conj{(zw)}=\conj w\conj z\;,\qquad\adj{(zw)}=\adj z\adj w\;.
\end{equation}

What appears to be a dual role of the conjugate in \cref{def:conj} is in fact one operation. 
In an anti-commutative basis, where $\e\alpha\e\beta=-\e\beta\e\alpha$, products behave as imaginary quantities. That is, if $\e\alpha$ behaves like a real quantity of unit length, then $(\e\alpha)^{-1}=\e\alpha$. The product of two such quantities has an inverse $(\e\alpha\e\beta)^{-1}=\e\beta\e\alpha$ (such that $(\e\alpha\e\beta)(\e\alpha\e\beta)^{-1}=\e\alpha\e\beta\e\beta\e\alpha=1$), but this is $(\e\alpha\e\beta)^{-1}=\e\beta\e\alpha=-\e\alpha\e\beta$ due to anti-commutivity. This implies $(\e\alpha\e\beta)^2=-1$. We may therefore write such a quantity as $\e\alpha\e\beta=\im\v u$ where $\v u$ is a real unit length vector. The action of the conjugate is then $\conj{(\im\v u)}=\conj{(\e\alpha\e\beta)}=-\e\alpha\e\beta=-\im\v u$. 

These operations allow us to decompose any quantity $z$ into constituent parts --- scalar and vector, real and imaginary --- by summing or subtracting the mirror, conjugate, or adjoint of $z$. 

It will sometimes be useful to denote components of the mirror, conjugate, or adjoint, of a (co)vector by applying the operator symbol to the index. So, for example, although the mirror only acts on the bases, the mirror of the vector $z$ can be written as
\begin{equation}\label{mirrind}
\mirror z=z^\alpha\mirror{\e\alpha}=z^{\mirror\alpha}\e\alpha\;.
\end{equation}

These operations are best extended to general bases by re-defining them in terms of tensor operators, say
\begin{equation}
\mirror{\e\alpha}=\op M^\beta_\alpha\e\beta\;.
\end{equation}
They can then be seen to be covariant, as a coordinate transformation $\Lambda$ commutes with the mirror operation $\op M$. So the mirror of $\e{\alpha'}$ taken in a transformed (primed) basis is 
\begin{align}
 \mirror{\e{\alpha'}}
& =\op M_{\alpha'}^{\beta'}\e{\beta'}\nonumber\\
& =(\Lambda_{\alpha'}^\mu\Lambda^{\beta'}_\nu\op M_\mu^\nu)(\Lambda_{\beta'}^\beta\e\beta)\nonumber\\
& =\Lambda_{\alpha'}^\mu\op M_\mu^\nu1_\nu^\beta\e\beta
 =\Lambda_{\alpha'}^\mu\op M_\mu^\nu\e\nu
\end{align}
which is the transformation of the mirror of $\e\alpha$ in the untransformed basis.

\subsection{The metric and mixture}\label{sec:metric}

The magnitude of a quantity $z$, which may be complex valued, is given by
\begin{align}\label{abs}
|z|^2&=z\mirror z=\mirror{z}z
=z^\alpha z^\beta\;\e\alpha\mirror{\e\beta}=z^\alpha z^\beta\eta_{(\alpha\mirror\beta)}^\gamma\e\gamma
\;,
\end{align}
with the round brackets denoting that the expression is symmetric in $\alpha\beta$. This imlpies that $\eta_{(\alpha\mirror\beta)}^\gamma\e\gamma$ should be a scalar. In the natural geometry in \cref{sec:natgeo}, for instance, this becomes $\eta_{(\alpha\mirror\beta)}^\gamma\e\gamma=\eta_{(\alpha\mirror\beta)}^0\e0$ where $\e0$ is the sole commuting basis. 

This magnitude traditionally defines the metric tensor $g$,
\begin{align}\label{absg}
|z|^2=z^\alpha z^\beta g_{\alpha\beta}=z^\alpha z_\alpha\;.\qquad\qquad\qquad\qquad
\end{align}
In the last equality of \cref{absg} we also introduce index lowering on a component $z^\beta$ via the metric. 

This implies that the metric is actually part of the mixture, given by 
\begin{equation}\label{glow}
g_{\alpha\beta}=\eta^\gamma_{(\alpha\mirror\beta)}\e\gamma
=\hf(\eta_{\alpha\mirror\beta}^\gamma+\eta_{\beta\mirror\alpha}^\gamma)\e\gamma
=\hf(\e\alpha\;\mirror{\e\beta}+\e\beta\;\mirror{\e\alpha})\;.
\end{equation}
Correspondingly for upper indices we define
\begin{equation}\label{ghigh}
g^{\alpha\beta}=\eta_\gamma^{(\alpha\mirror\beta)}\ec\gamma
=\hf(\eta^{\alpha\mirror\beta}_\gamma+\eta^{\beta\mirror\alpha}_\gamma)\ec\gamma
=\hf(\ec\alpha\;\mirror{\ec\beta}+\ec\beta\;\mirror{\ec\alpha})\;.\;
\end{equation}

\subsection{A natural geometry}\label{sec:natgeo}

As an example algebra let us take the geometry of everyday mechanics, namely that of displacements and rotations. 
Let $\e1,\e2,\e3$, be the orthogonal vector bases in physical space, and $\e0$ a scalar basis. The spatial bases are anti-commutative. 

Let these bases be real, so that they are their own inverses, $1/\e\alpha=\e\alpha$ for $\alpha=0,1,2,3$, but let them be anti-commutative such that $\e i\e j=\e j\e j$ if $i\neq j$ for $i,j=1,2,3$. The product $\e i\e j$ cannot lie in the plane of $\e i$ and $\e j$ but must instead lie along the third spatial direction $\e k$, and yet must be imaginary since $(\e i\e j)^2=\e i\e j\e i\e j=-\e i(\e j)^2\e i=-1$. We then have (as set out more fully in \cref{sec:Anatgeo}) the multiplication rules
\begin{align}\label{hamilton}
\;\qquad(\e\alpha)^2=\e0\;\;\qquad{\rm and}\qquad\e i\e j=-\e j\e i=\im\e k\;,
\end{align}
where and $\im=${\small$\sqrt{-1}$}, with $\cc{i,j,k}$ being cyclic permutations of the indices $\cc{1,2,3}$. Thus while $\e i$ denotes a vector, the imaginary $\im\e i$ denotes a rotation about that vector, for instance rotating $\e j$ to $\e k$ since $(\im\e i)\e j=\im^2\e k=-\e k$ (where again $\cc{i,j,k}$ are a cyclic permutation of $\cc{1,2,3}$).

These are essentially just an extension to Hamilton's quaternion rules that distinguishes vectors from rotations by use of the imaginary (in fact Hamilton's quaternion bases \cite{H1846} are $\e0,\im\e1,\im\e2,\im\e3$). To rotate the vector $\e1$ by angle $\omega$ about the direction $\e3$, for example, we write
\begin{align}
\e1\exp{\im\e3\omega}=\e1(\e0\cos\omega+\im\e3\sin\omega)=\e1\cos\omega+\e2\sin\omega\;.
\end{align}
As is well known, one-sided multiplication like $z\exp{\im\e3\omega}$ produces a `double rotation' in the planes of $\e0$-$\e3$ and $\e1$-$\e2$; to extract purely the spatial rotation in the $\e1$-$\e2$ plane we multiply as $\exp{-\hf\im\e3\omega}z\exp{\hf\im\e3\omega}$, see e.g. \cite{geoma}. 



%
\begin{figure}[h!]\centering\includegraphics[width=0.6\textwidth]{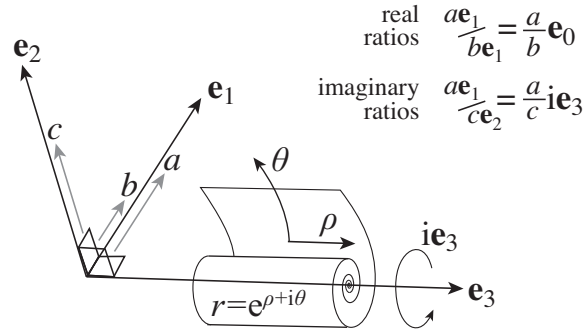}
\vspace{-0.3cm}\caption{\footnotesize\sf Basis vectors $\e\alpha$, rotations $\ec\alpha$, and the duality via the imaginary $\im=\sqrt{-1}$. }\label{fig:quat}\end{figure}

This self-contained algebra, in which complex numbers provide rotations about vector axes, should be considered as a separate framework to the geometric algebras which consider exterior products between vectors to form multi-vectors associated with areas, volumes, etc.. 

We will call this the {\it natural geometry}, and can denote it alternatively by $\mathbb C^{1+3}=\mathbb C\times\mathbb H=\mathbb R^{1+3}+\mathbb I^{1+3}$ or other such forms, where $\mathbb R,\mathbb I,\mathbb C,\mathbb H,$ denote real, imaginary, compex, and quaternion Euclidean space; for shorthand let us denote $\mathbb C^{1+3}$ as $\rasps$. It can be used as a basis for most investigations in this article, until we come to deal with specific differentiability conditions such as, for example, the Klein-Gordon equations. 

In the natural geometry, the mixture encodes the relations \cref{hamilton}, and its components are
\begin{align}\label{etaflat}
&\eta^0_{\alpha\alpha}=\eta^\alpha_{0\alpha}=\eta^\alpha_{\alpha0}=1\;,\qquad\eta^i_{jk}=-\eta^i_{kj}=\im\;,\\
&\eta^i_{\alpha\alpha}=\eta^\alpha_{i\alpha}=\eta^\alpha_{\alpha i}=0\;,\nonumber
\end{align}
(not summing over indices), where again $\cc{ijk}$ are cyclic permutations of $\cc{123}$. The lower and upper index metric tensors $g_{\alpha\beta}$ and $g^{\alpha\beta}$ as defined by \cref{glow}-\cref{ghigh} have diagonal components $\cc{+1,-1,-1,-1}$ in this algebra.

\begin{lemma}\label{thm:etainv}
The (pseudo-)inverse of $\eta_{\alpha\beta}^\gamma$ is $\sfrac14\eta^{\beta\alpha}_\gamma$, such that
\begin{equation}\label{etainverse}
\eta_{\beta\gamma}^\alpha\eta^{\gamma\beta}_\delta=
\eta_{\delta\beta}^\gamma\eta^{\beta\alpha}_\gamma=
\eta^{\alpha\beta}_\gamma\eta_{\beta\delta}^\gamma=n1^\alpha_\delta\;,
\end{equation}
where $1^\alpha_\delta$ denotes the identity matrix. 
\end{lemma}
\vspace{-0.2cm}\noindent Proof: To show that $\eta_{\beta\gamma}^\alpha\eta^{\gamma\beta}_\delta=n1^\alpha_\delta$, multiply by $\e\alpha$, then by \cref{eta} we have $\eta_{\beta\gamma}^\alpha\eta^{\gamma\beta}_\delta\e\alpha=\e\beta\e\gamma\eta^{\gamma\beta}_\delta=\e\beta\ec\beta\e\delta=n\e\delta=n1^\alpha_\delta\e\alpha$. Comparing the first and last terms for any $\delta$ gives the result. \bigskip

In addition the mixture satisfies several relations concerning index exchanges and symmetries, some of which are outlined in \cref{sec:Amixture}.

Multiplying the two parts of \cref{eta} by $\ec\delta$ and $\e\delta$ respectively, then adding the mirror of each expression, gives
\begin{align*}
&\e\alpha \e\beta\ec\delta+\mirror{\e\alpha \e\beta\ec\delta}=\eta^\gamma_{\alpha\beta}\eta_\gamma^{\delta\lambda}(\e\lambda+\mirror{\e\lambda})=2\eta^\gamma_{\alpha\beta}\eta_\gamma^{\delta0}\e0=2\eta^\delta_{\alpha\beta}\\
&\ec\alpha \ec\beta\e\delta+\mirror{\ec\alpha \ec\beta\e\delta}=\eta_\gamma^{\alpha\beta}\eta^\gamma_{\delta\lambda}(\ec\lambda+\mirror{\ec\lambda})=2\eta_\gamma^{\alpha\beta}\eta^\gamma_{\delta0}\ec0=2\eta_\delta^{\alpha\beta}
\end{align*}
giving formulae for the mixture in terms of the bases, 
\begin{align}
\eta^\delta_{\alpha\beta}=\sfrac1{2}\bb{\e\alpha \e\beta\ec\delta+\mirror{\e\alpha \e\beta\ec\delta}}\;,\quad
\eta_\delta^{\alpha\beta}=\sfrac1{2}\bb{\ec\alpha \ec\beta\e\delta+\mirror{\ec\alpha \ec\beta\e\delta}}\;.
\end{align}

In curved bases these relations typically no longer hold exactly, but $\eta$ transforms as a tensor, as we shall see in \cref{sec:covariance}. 

As a generalization of the quaternions this algebra has various nice geometrical characteristics, and it is perhaps worth laying out a little of its behaviour.  

\subsection{Zeros and exponentials in the natural geometry}


An important feature of the natural geometry is its null quantities. 

Null quantities are those $\hat n$ such that $|\hat n|=0$, and are indicative of the hyperbolic part of the natural geometry. Note that while the square of $\e1$ is $\e1^2=1$, its length is $|\e1|^2=\e1\mirror{\e1}=-1$. This implies that if ${\bf x}\in\mathbb R^3$ and $\hat {\bf x}={\bf x}/|{\bf x}|$, then $|{\bf x}|^2=-{\bf x}^2$, and $\hat n=1\pm\im\hat{\bf x}$ is a null element since 
\begin{equation}
\left|1\pm\im\hat{\bf x}\right|^2=(1\pm\im\hat{\bf x})(1\mp\im\hat{\bf x})=1+\hat{\bf x}^2=1+{\bf x}^2/|{\bf x}|^2=1-1=0\;.
\end{equation}
For example $1+\e1$ is null, while the quaternion $1+\im\e1$ has length $\sqrt2$. 

If some $\hat n$ is null, then given any other $\z$ (null or otherwise), the products $\z\hat n$ and $\hat n\z$ are null since
\begin{equation}
|\z\hat n|=|\hat n\z|=|\z||\hat n|=0\;.
\end{equation}
We may refer to this as the {\it confinement of null quantities} --- they are confined to remaining null under multiplication and division, and hence behave rather like zero.


A point where some quantity $q$ vanishes is not generally isolated, but may generate a null set on which $|q|=0$. In the natural geometry, letting $q=t+\v x=t\e0+x\e1+y\e2+z\e3$ (where $t,x,y,z,$ may themselves be complex), the null set about $q=0$ is the diabolical (or bi-conical) hypersurface
\begin{align}
|q|^2=t^2-\v x^2=t^2-x^2-y^2-z^2=0\;.
\end{align}

Consider now writing a function in the form $g=\exp{\phi}$. This allows us to extract certain geometry by breaking up the exponent into real and imaginary, scalar and vector, parts, 
\begin{align}\label{ephigeo}
\exp{\phi}
&=\exp{\alpha+\im\beta+\v c+\im\v d}\nonumber\\
&=\exp{\alpha+\im\beta+(\gamma+\im\delta)\v k}\nonumber\\
&=\exp{\alpha+\gamma\v k}\exp{\im(\beta+\delta\v k)}
\end{align}
where $\v k^2=1$, as $\v k=\im\frac{\v c+\im\v d}{|\v c+\im\v d|}$ and $\gamma+\im\delta=|\v c+\im\v d|=\sqrt{\v d^2-\v c^2-2\im\v c\cdot\v d}$. 
The last line of \cref{ephigeo} splits $\exp\phi$ into its 
evanescent part $|\exp\phi|=\exp{\alpha+\gamma\v k}$, and its oscillatory part $\exp{\im(\beta+\delta\v k)}$ satisfying
\begin{align}
&|\exp{\im(\beta+\delta\v k)}|=1\quad\&\quad
\exp{\im(\beta+\delta\v k)}=\exp{\im(\beta+\delta\v k+2\pi n)}\;,\quad n\in\mathbb Z\;.
\end{align}


Consider now if $\phi$ is a function of $z$. 
Along a contour $z=z(s)$, parameterized by $s\in\reals$, say that $\phi$ at some $z_0$ has a value
\begin{align}\label{phi0}
\phi(z_0):=\;\left.\phi\big(z(s)\big)\right|_{s=0}\;=\alpha+\im\beta+(\gamma+\im\delta)\v k\quad
\end{align}
and its derivative along the contour there is
\begin{align}\label{dphi0}
\phi'(z_0):=\left.\sfrac{d\;}{ds}\phi(z(s))\right|_{s=0}=\xi+\im\eta+\v p+\im\v q\;.\qquad
\end{align}
The behaviour of $\exp\phi$ along $z(s)$ is rather complicated if $\v p$ and $\v q$ do not commute with $\v k$. 
So imagine that we can change the path of $z(s)$ such that
\begin{align}\label{dphi0d}
\phi'(z_0)=\left.\sfrac{d\;}{ds}\phi(z(s))\right|_{s=0}=\xi+\im\eta+(\zeta+\im\omega)\v k\;.\qquad
\end{align}
Then
\begin{align}\label{ephi}
\exp{\phi(z(s))}=\exp{\alpha+\im\beta+(\xi+\im\eta)s}\exp{[\gamma+\im\delta+(\zeta+\im\omega)\v k]s+\ord{s^2}}
\end{align}
and the vector part of $\phi'(z_0)$ commutes with the vector part of $\phi(z_0)$, hence they produce no oscillation that would results from a cross-product term. 

If we moreover deform such that 
\begin{align}\label{dphi0d0i}
\imp{\left.\sfrac{d\;}{ds}\phi(z(s))\right|_{s=0}}=\eta+\omega\v k=0
\end{align}
(we shall use $\rep{..}$ and $\imp{..}$ to denote the real and imaginary parts of quantities), 
then there is no oscillation, and only steepest descent, along the path $z(s)$, locally at least. 
If we deform such that 
\begin{align}\label{dphi0d0r}
\rep{\left.\sfrac{d\;}{ds}\phi(z(s))\right|_{s=0}}=\xi+\zeta\v k=0
\end{align}
then $\exp\phi$ is purely oscillating along the path $z(s)$, without any decay, again locally at least. 

Such deformation of paths is necessary to find a path that minimizes an integral, or to solve an integral by steepest descent or stationary phase methods. This relies on such contour deformation being permitted (i.e. the integrand being integrable), and on a path through $z_0$ existing satisfying \cref{dphi0d0i} or \cref{dphi0d0r}. 


Consider a couple of examples. 

\medskip

\begin{example}
If $\phi=\hf (u-i z)^2=\hf (u^2-z^2)- z\cdot u$, then \cref{dphi0d0r} implies
\begin{align}\label{eikham}
0=\rep{\left.\sfrac{d\;}{ds}\phi(z(s))\right|_{s=0}}
=\hf\sfrac{d\;}{ds}(u^2-z^2)=u\cdot\dot u-z\cdot\dot z\;,
\end{align}
the obvious solution of which is $\dot z=u$, $\dot u=-z$, defining a Hamiltonian system on the space-time vector $z=ct\e0+x^i\e i$ and 4-momentum vector $u=\sfrac Ec\e0+p^i\e i$. 
\end{example}

\begin{example}
Now consider if $\phi(z(s))=u$, and $u=\dot z$. Denote the (local) derivative with respect to $s$ with a dot, then \cref{dphi0d0r} implies
\begin{align}\label{geodesiC}
0=\rep{\left.\sfrac{d\;}{ds}\phi(z(s))\right|_{s=0}}
=\rep{\dot u}=\rep{\dot z^\beta\sfrac{d\;}{dz^\beta}u}=\rep{u^\beta u^\alpha_{;\beta}}\e\alpha\;.
\end{align}
If $u$ is real then from the zero descent condition \cref{dphi0d0r} we have the parallel transport (or geodesic) condition $u^\beta u^\alpha_{;\beta}=0$. 
\end{example}

We will use \cref{geodesiC} to extend the notion of parallel transport to complex vectors, that is, we say the path $z(s)$ parallel transports its own tangent vector $u=\dot z$ if the real part of $u$ remains parallel to $z(s)$, while the imaginary part may wander; we can picture this as $u$ being able to rotate (in the complex plane) while remaining parallel (with regard to the vector bases) to its path.  

\section{Covariant calculus (with mixtures)}\label{sec:covariance}

The mixture offers an alternative to previous means of handling geometric algebras, 
but it also lends itself very naturally to the covariant calculus over curved manifolds as largely formalized by Christoffel and Levi-Civitia \cite{levicivita}. We must make modifications to base the theory around the mixture rather than the metric which, while seemingly slight, have profound implications.

\subsection{The connection: variation of bases}

Let $\Lambda_\alpha^{\alpha'}$ be the transformation matrix from a constant (Euclidean) basis $\e{\alpha'}$ into a varying basis $\e\alpha=\Lambda_\alpha^{\alpha'}\e{\alpha'}$, and let $\Lambda^\alpha_{\alpha'}$ denote its inverse. 

Denote a partial derivative as $\sfrac{\partial\;}{\partial z^{\beta}}w\equiv w_{,\beta}$ for any $w$.  
The derivative of a basis $\e\alpha$ or dual $\ec\alpha$ can be written as
\begin{align}\label{edcon}
\e{\alpha,\beta}=\con^{\gamma}_{\alpha\beta}\e{\gamma}\;,\qquad
\ec\alpha_{,\beta}=-\con^{\alpha}_{\gamma\beta}\ec{\gamma}\;,
\end{align}
in terms of the Christoffel symbols $\con^\gamma_{\alpha\beta}$. In a slight but convenient abuse of terminology we will call $\con$ itself the {\it connection} symbol. Calculating the full derivative $\sfrac{d\;}{dz}=\ec\beta\sfrac{\partial\;}{\partial z^{\beta}}$, 
\begin{align}\label{edconcharge}
\ec\beta\e{\alpha,\beta}=W^\delta_\alpha\e\delta\;,\qquad
\ec\beta\ec\alpha_{,\beta}={\raisebox{\depth}{\rotatebox{180}{\mbox{$W$}}}}^\alpha_\delta\ec\delta\;,
\end{align}
where $W^\delta_\alpha=\con^{\gamma}_{\alpha\beta}\eta^{\delta\beta}_\gamma$ and ${\raisebox{\depth}{\rotatebox{180}{\mbox{$W$}}}}^\alpha_\delta=-\con^{\alpha}_{\gamma\beta}\eta^{\beta\gamma}_\delta$. 

Differentiating a basis twice gives
\begin{align}\label{fR}
\ec\nu\sfrac{\partial\;}{\partial w^{\nu}}\ec\mu\sfrac{\partial\;}{\partial w^{\mu}}\e{\beta}=\ec\nu\ec\mu\e{\beta,{\mu\nu}}
&=(\con^{\alpha}_{\sigma\nu}\con^{\sigma}_{\beta\mu}+\con^{\alpha}_{\beta\mu,\nu}
)\ec\nu\ec\mu\e{\alpha}\nonumber\\
:&=\scon^{\alpha}_{\beta\nu\mu}\ec\nu\ec\mu\e{\alpha}\;
\end{align}
defining a {\it second connection} symbol $\scon$. 
Antisymmetrizing over the $\mu\nu$ indices gives the Riemann curvature tensor, 
\begin{align}\label{fR}
R^{\alpha}_{\beta\nu\mu}:&=P^\alpha_{\beta[\nu\mu]}\\\nonumber&=\con^{\alpha}_{\nu\sigma}\con^{\sigma}_{\mu\beta}+\con^{\alpha}_{\mu\beta,\nu}-\con^{\alpha}_{\mu\sigma}\con^{\sigma}_{\nu\beta}-\con^{\alpha}_{\nu\beta,\mu}
\;.
\end{align}

The Riemann tensor is known to have $n^2(n^2-1)/12$ free components, which in $n=4$ dimensions amount to 16 components plus 4 coordinate freedoms. We could therefore introduce a tensor $K_{\gamma\beta}$ with $n^2$ free components, and 
try a particular ansatz $R^{\alpha}_{\gamma\nu\mu}=K_{\gamma\beta}\eta_{\lambda}^{\alpha\beta}\eta^{\lambda}_{\mu\nu}$. Taking the trace gives the Ricci tensor and Ricci scalar as
\begin{align}
%
R_{\gamma\mu}&=R^{\alpha}_{\gamma\alpha\mu}=
K_{\gamma\beta}\eta_{\lambda}^{\alpha\beta}\eta^{\lambda}_{\mu\alpha}=
4 K_{\gamma\mu}\nonumber\\
\Rightarrow\qquad\quad
R&=g^{\gamma\mu}R_{\gamma\mu}=
4 K\;.
\end{align}
For this ansatz the antisymmetry of $R^{\alpha}_{\gamma\nu\mu}$ implies
\begin{align}
0=K_{\gamma\beta}\eta_{\lambda}^{\alpha\beta}\eta^{\lambda}_{(\mu\nu)}\;.
\end{align}
\subsection{Covariant differentiation of components}

As in standard theory, the covariant derivative of a vector $f=f^\alpha\e\alpha$ is 
\begin{align}
\sfrac{d\;}{d w^{\beta}}f=f^{\alpha}_{;\beta}\e{\alpha}\qquad\;\;\;{\rm where}\quad 
f^{\alpha}_{;\beta}&=f^{\alpha}_{,\beta}+f^{\gamma}\con^{\alpha}_{\gamma\beta}\;,\;\;
\end{align}
while for a dual vector it is
\begin{align}
\sfrac{d\;}{d w^{\beta}}\tilde f=f_{\alpha;\beta}\e{\alpha}\qquad{\rm where}\quad 
f_{\alpha;\beta}&=f_{\alpha,\beta}-f_{\gamma}\con^{\gamma}_{\alpha\beta}\;.
\end{align}

To differentiate the mirror or conjugate of a vector is just as simple. Although the mirror is defined as acting on a basis, if we use the index notation from \cref{mirrind} in which the component $f^{\mirror\alpha}$ of the mirror $\mirror f$ can be treated like any vector component, we see that 
$$\sfrac{d\;}{d w^{\beta}}f=\mirror{f}_{;\beta}=f^{\mirror \alpha}_{,\beta}\e{\alpha} \;=\;(f^{\mirror\alpha}_{,\beta}+f^{\mirror\gamma}\con^{\alpha}_{\gamma\beta})\e{\alpha}\;.$$

For the second derivative of $f$ we have
\begin{align*}
%
\quad\;\;
 f^{\alpha}_{;\mu\nu}\ec\nu\ec\mu\e{\alpha}&=\ec\nu\cc{(f^{\alpha}_{,\mu}+f^{\gamma}\con^{\alpha}_{\gamma\mu})\ec\mu\e{\alpha}}_{;\nu}\nonumber\\
&=\left\{
f^{\gamma}(\con^{\lambda}_{\gamma\mu}\con_{\lambda\nu}^\alpha+\con^{\alpha}_{\gamma\mu,\nu}-\con^{\alpha}_{\gamma\lambda}\con_{\mu\nu}^\lambda)\right.\nonumber\\&\qquad+\left.
(f^{\gamma}_{,\mu}\con_{\gamma\nu}^\alpha+f^{\gamma}_{,\nu}\con^{\alpha}_{\gamma\mu}-f^{\alpha}_{,\gamma}\con_{\mu\nu}^\gamma)+
f^{\alpha}_{,\mu\nu}
\right\}\ec\nu\ec\mu\e\alpha\nonumber\\
&=\left\{
f^{\gamma}(\scon^{\alpha}_{\gamma\nu\mu}-\con^{\alpha}_{\gamma\lambda}\con_{\mu\nu}^\lambda)
\right.\nonumber\\&\qquad\left.  +  (2\con_{\gamma(\nu}^\alpha f^{\gamma}_{,\mu)}-f^{\alpha}_{,\gamma}\con_{\mu\nu}^\gamma)+
f^{\alpha}_{,\mu\nu}  \right\}\ec\nu\ec\mu\e\alpha\nonumber\\
\Rightarrow\qquad 
%
%
 f^{\alpha}_{;\mu\nu}\eta^{\nu\mu}_{\delta}\eta^{\delta}_{\alpha\omega}
&=\left\{
f^{\gamma}(\scon^{\alpha}_{\gamma\nu\mu}-\con^{\alpha}_{\gamma\lambda}\con_{\mu\nu}^\lambda)
\right.\nonumber\\&\qquad\left.  
+  (2\con_{\gamma(\nu}^\alpha f^{\gamma}_{,\mu)}-f^{\alpha}_{,\gamma}\con_{\mu\nu}^\gamma)+
f^{\alpha}_{,\mu\nu} \right\}\eta^{\nu\mu}_{\delta}\eta^{\delta}_{\alpha\omega}\\
\Rightarrow\qquad \!\!\!
 f^{\alpha}_{;[\mu\nu]}\eta^{\nu\mu}_{\delta}\eta^{\delta}_{\alpha\omega}
&=\cc{
f^{\gamma}R^{\alpha}_{\gamma\nu\mu}
-f^{\alpha}_{;\gamma}\con_{[\mu\nu]}^\gamma }\eta^{\nu\mu}_{\delta}\eta^{\delta}_{\alpha\omega}
\end{align*}
%
%
%
giving a second order differential equation relating $f$ to $\con$ and $R$, 
\begin{align}
0
&=\cc{ f^{\alpha}_{;[\mu\nu]}+f^{\alpha}_{;\gamma}\con_{[\mu\nu]}^\gamma-f^{\gamma}R^{\alpha}_{\gamma\nu\mu}
 }\eta^{\nu\mu}_{\delta}\eta^{\delta}_{\alpha\omega}\;.
\end{align}
\subsection{Connection and metric}

To verify that the connection as derived above is indded the familiar Christoffel symbol, and also to motivate an expression relating the connection and mixture in the next section, let us differentiate $g_{\alpha\beta}=\e\alpha\cdot\e{\mirror\beta}$ using the definition $\e{\alpha,\beta}=\con^\gamma_{\alpha\beta}\e\gamma$, giving
\begin{align}
g_{\alpha\beta,\mu}&=\e{\alpha,\mu}\cdot\e{\mirror\beta}+\e\alpha\cdot\e{\mirror\beta,\mu}\nonumber\\
&=\con^\gamma_{\alpha\mu}\e\gamma\cdot\e{\mirror\beta}+\con^{\mirror\gamma}_{\mirror\beta\mu}\e\alpha\cdot\e{\mirror\gamma}=\con^\gamma_{\alpha\mu}g_{\gamma\beta}+\con^{\gamma}_{\beta\mu}g_{\alpha\gamma}\;,
\end{align}
which implies  
\begin{equation}\label{met0}
g_{\alpha\beta;\mu}=g_{\alpha\beta,\mu}-\con^\gamma_{\alpha\mu}g_{\gamma\beta}-\con^{\gamma}_{\beta\mu}g_{\alpha\gamma}=0\;.
\end{equation}
If we permute the indices we obtain
\begin{align}
0&=g_{\alpha\beta,\mu}-g_{\lambda\beta}\con^\lambda_{\alpha\mu}-g_{\alpha\lambda}\con^\lambda_{\beta\mu}\nonumber\\
&=g_{\beta\mu,\alpha}-g_{\lambda\mu}\con^\lambda_{\beta\alpha}-g_{\beta\lambda}\con^\lambda_{\mu\alpha}\nonumber\\
&=g_{\mu\alpha,\beta}-g_{\lambda\alpha}\con^\lambda_{\mu\beta}-g_{\mu\lambda}\con^\lambda_{\alpha\beta}\;,
\end{align}
then summing the first two lines and subtracting the third, exploiting the symmetry of $g$ over its indices, we have 
\begin{align*}
0&=g_{\alpha\beta,\mu}+g_{\beta\mu,\alpha}-g_{\mu\alpha,\beta}\\&\qquad
-2g_{\lambda\beta}\con^\lambda_{(\alpha\mu)}
-2g_{\alpha\lambda}\con^\lambda_{[\beta\mu]}
-2g_{\lambda\mu}\con^\lambda_{[\beta\alpha]}\;.
\end{align*}
Multiplying by $g^{\beta\sigma}$ gives 
\begin{align*}
0&=\hf g^{\beta\sigma}(g_{\alpha\beta,\mu}+g_{\beta\mu,\alpha}-g_{\mu\alpha,\beta})\\&\qquad
-\con^\sigma_{(\alpha\mu)}
-g^{\beta\sigma}(g_{\alpha\lambda}\con^\lambda_{[\beta\mu]}+g_{\lambda\mu}\con^\lambda_{[\beta\alpha]})\;.\quad
\end{align*}
which can be re-arranged to
\begin{align}\label{cong}
\con^\sigma_{\alpha\mu}&=\hf g^{\beta\sigma}(g_{\alpha\beta,\mu}+g_{\beta\mu,\alpha}-g_{\mu\alpha,\beta})
\\\nonumber&\qquad-g^{\beta\sigma}(g_{\lambda\mu}\con^\lambda_{[\beta\alpha]}+g_{\alpha\lambda}\con^\lambda_{[\beta\mu]}-g_{\beta\lambda}\con^\lambda_{[\alpha\mu]})\;.
\end{align}
This has the familiar symmetric part
\begin{equation}\label{christoffel}
\cris^\sigma_{(\alpha\mu)}=\hf g^{\beta\sigma}(g_{\alpha\beta,\mu}+g_{\beta\mu,\alpha}-g_{\mu\alpha,\beta})\;,\qquad\qquad\quad\;\;
\end{equation}
so in full we can write
\begin{equation}
\con^\sigma_{\alpha\mu}=\cris^\sigma_{(\alpha\mu)}
+\hf g^{\beta\sigma}(C_{\beta\alpha\mu}+C_{\beta\mu\alpha}-C_{\alpha\mu\beta})\;\qquad
\end{equation}
in term of commutation coefficients $C_{\beta\alpha\mu}=-2g_{\lambda\mu}\con^\lambda_{[\beta\alpha]}$. 
If we raise the index of $C$ we have
\begin{equation}
C_{\beta\alpha}^\delta=g^{\mu\delta}C_{\beta\alpha\mu}=-2g^{\mu\delta}g_{\lambda\mu}\con^\lambda_{[\beta\alpha]}=2\con^\delta_{[\alpha\beta]}\;,
\end{equation}
which are Cartan's commutation coefficients \cite{cartan1922,misner73} (these are given, if we consider a vector field to the generator of a flow such that $\e\alpha=\sfrac{\partial\;}{\partial z^\alpha}$, by the Lie bracket $c_{\beta\alpha}^\delta\e\delta=L\sq{\e\beta,\e\alpha}=\sfrac{\partial\;}{\partial z^\alpha}\e\beta-\sfrac{\partial\;}{\partial z^\beta}\e\alpha$, which in our formalism reads $c_{\beta\alpha}^\delta\e\delta=2\con_{[\alpha\beta]}^\delta\e\delta$). 
So the antisymmetric part of the connection $\con$ is given by Cartan's commutation coefficients, but the symmetric part of the connection also in general involves a contribution from the commutation coefficients, 
\begin{align}
\con^\sigma_{(\alpha\mu)}=\cris^\sigma_{(\alpha\mu)}
+ g^{\beta\sigma}C_{\beta(\alpha\mu)}\;,\qquad
\con^\sigma_{[\alpha\mu]}=\hf g^{\beta\sigma}C_{\mu\alpha\beta}=\hf C^\sigma_{\mu\alpha}\;.
\end{align}

%
%

\bigskip

We should be able to derive the result \cref{met0} also from the mixture. Differentiating the mixture with the mirror on one index gives 
\begin{equation}
\eta_{\alpha\mirror\beta;\mu}^\gamma=\eta_{\alpha\mirror\beta,\mu}^\gamma+\eta_{\alpha\mirror\beta}^\lambda\con_{\lambda\mu}^\gamma-\eta_{\lambda\mirror\beta}^\gamma\con_{\alpha\mu}^\lambda-\eta_{\alpha\mirror\lambda}^\gamma\con_{\beta\mu}^{\lambda}\;.
\end{equation}
Symmetrizing over the lower indices, and multiplying by $\e\gamma$, if the only commuting basis is $\e0$, we have
\begin{align}
\eta_{(\alpha\mirror\beta);\mu}^0\e0&=\bb{\eta_{(\alpha\mirror\beta),\mu}^\gamma+\eta_{(\alpha\mirror\beta)}^0\con_{0\mu}^\gamma}\e\gamma-\eta_{(\lambda\mirror\beta)}^0\con_{\alpha\mu}^\lambda\e0-\eta_{(\alpha\mirror\lambda)}^0\con_{\beta\mu}^\lambda\e0\nonumber\\
\Rightarrow\qquad g_{\alpha\beta;\mu}&=g_{\alpha\beta,\mu}-g_{\lambda\beta}\con_{\alpha\mu}^\lambda-g_{\alpha\lambda}\con_{\beta\mu}^\lambda
\end{align}
in agreement with \cref{met0}. The implication in the second line makes use of $g_{\lambda\beta}=\eta_{(\lambda\mirror\beta)}^0\e0$ for the last two terms, of $g_{\alpha\beta;\mu}=\eta_{\alpha\mirror\beta;\mu}^0\e0$ for the lefthand side, and less obviously for the term $g_{\alpha\beta,\mu}$ of the fact that $(g_{\alpha\beta})_{,\mu}=(\eta_{(\alpha\mirror\beta)}^0\e0)_{,\mu}=(\eta_{(\alpha\mirror\beta),\mu}^\gamma+\eta_{(\alpha\mirror\beta)}^0\con_{0\mu}^\gamma)\e\gamma$. This last relation must be treated with care, and implies
\begin{align}\label{gimportant}
g_{\alpha\beta,\mu}&=(\eta_{(\alpha\mirror\beta),\mu}^0+g_{\alpha\beta}\con_{0\mu}^0)\e0\;,\\
\&\qquad\qquad
0&=\eta_{(\alpha\mirror\beta),\mu}^i+g_{\alpha\beta}\con_{0\mu}^i\;,\qquad\mbox{\footnotesize$i=1,2,3,...$}\;.\nonumber
\end{align}

\subsection{Connection and mixture}

To relate $\con$ and $\eta$, emulating the derivation of the metric relation above, let us differentiate the full product product $\e\alpha\e\beta$ in two different ways. If we differentiate each element before mixing them we get
\begin{align}
(\e\alpha\e\beta)_{,\mu}&=\e{\alpha,\mu}\e\beta+\e\alpha\e{\beta,\mu}\nonumber\\
&=(\con_{\alpha\mu}^\lambda\eta_{\lambda\beta}^\gamma+\con_{\beta\mu}^\lambda\eta_{\alpha\lambda}^\gamma)\e\gamma\qquad\qquad\;\;\;
\end{align}
and if we mix them before differentiating we get
\begin{align}
(\e\alpha\e\beta)_{,\mu}=(\eta_{\alpha\beta}^\gamma\e\gamma)_{,\mu}&=\eta_{\alpha\beta,\mu}^\gamma\e\gamma+\eta_{\alpha\beta}^\gamma\e{\gamma,\mu}\nonumber\\&=(\eta_{\alpha\beta,\mu}^\gamma+\eta_{\alpha\beta}^\lambda\con_{\lambda\mu}^\gamma)\e\gamma\;.
\end{align}
Equating the two expressions gives
\begin{align}\label{etadvanish}
0=\eta_{\alpha\beta,\mu}^\gamma+\eta_{\alpha\beta}^\lambda\con_{\lambda\mu}^\gamma-\eta_{\lambda\beta}^\gamma\con_{\alpha\mu}^\lambda-\eta_{\alpha\lambda}^\gamma\con_{\beta\mu}^\lambda=\eta_{\alpha\beta;\mu}^\gamma\;.
\end{align}
Thus the covariant derivative of the mixture, like that of the metric, vanishes. 

We can re-write this last relation as
\begin{align}\label{etageodesic}
0=\eta_{\alpha\beta,\mu}^\gamma+(1^\gamma_\delta\eta_{\alpha\beta}^\lambda-1^\lambda_\alpha\eta_{\delta\beta}^\gamma-1^\lambda_\beta\eta_{\alpha\delta}^\gamma)\con_{\lambda\mu}^\delta\;,
\end{align}
implying
\begin{align}
\con_{\lambda\mu}^\delta&=W^{\delta\alpha\beta}_{\gamma\lambda}\eta_{\alpha\beta,\mu}^\gamma
\end{align}
where
\begin{align}
W^{\delta\alpha\beta}_{\gamma\lambda}=[1^\gamma_\delta\eta_{\alpha\beta}^\lambda-1^\lambda_\alpha\eta_{\delta\beta}^\gamma-1^\lambda_\beta\eta_{\alpha\delta}^\gamma]^{-1}\;
\end{align}
Multiplying \cref{etageodesic} by $\eta$ gives
\begin{align}
0&=\eta_\gamma^{\delta\mu}\cc{\eta_{\alpha\beta,\mu}^\gamma+\eta_{\alpha\beta}^\lambda\con_{\lambda\mu}^\gamma-\eta_{\lambda\beta}^\gamma\con_{\alpha\mu}^\lambda-\eta_{\alpha\lambda}^\gamma\con_{\beta\mu}^\lambda}\nonumber\\
&=\eta_\gamma^{\delta\mu}\cc{\eta_{\alpha\beta,\mu}^\gamma+(1^\gamma_\lambda\eta_{\alpha\beta}^\omega-1^\omega_\alpha\eta_{\lambda\beta}^\gamma-1^\omega_\beta\eta_{\alpha\lambda}^\gamma)\con_{\omega\mu}^\lambda}\;,
\end{align}
a rank three tensor equation, which it seems should be invertible to find $\con$ as a function of $\eta$ and its derivatives, but as yet a solution evades the author. 


\section{Analytic curvature}\label{sec:curvature}

The analyticity of complex functions is a powerful tool for the calculation of scalar integrals, making possible contour deformations that pave the way for methods of residues and steepest descents. 
Extending these ideas wholesale from scalar variables to higher dimensions is made impossible by non-commutativity of bases. The definition of an analytic function as having a power series is useless in a non-commutative algebra, as such a power series is no longer unique, does not uniquely relate to dependence on a variable $z$ and any conjugates $\conj z$ (or $\mirror z$, $\adj z$, etc.), and is not clearly related to the vanishing of any derivative. Before we seek to extend conditions like the Cauchy-Riemann equations, for example, to non-commutative agebras, we first need to understand more what their significance is. We shall then see how they re-appear in a covariant theory as conditions for integrability and differentiability of functions.

So we wish to ask under what conditions we can integrate a function $g$ along some contour, schematically given by \cref{fund}, assuming a geometric product between $g$ and $dz$, such that $z,g$, and $f$ occupy the same algebraic space. 

We can strip \cref{fund} back a little by removing the integral sign to reveal the infinitesimal, or local, problem. Reading from right to left, \cref{fund} can then be interpreted as expanding $f$ in a (multivariable) Taylor series, 
\begin{align}\label{Taymin}
df&=\sfrac{df}{dz}dz+\ord{dz^2}=g\;dz+\ord{dz^2}\;.
\end{align}
Integrating over such increments yields the fundamental theorem \cref{fund}. 
Our problem becomes that of finding when such a series expansion \cref{Taymin} exists for $z,f,g\in\op S$ on some space $\op S$. 
This turns the problem from one of integrability to differentiability. 

It turns out, however, that only trivial functions are differentiable in this sense in higher dimensions. To take the form \cref{Taymin} places strong restrictions on $f$. These are satisfied by the Cauchy-Riemann equations for $f,z\in\mathbb \coms$, but in higher dimensions can only be satisfied if the derivative vanishes identically, i.e. by constant functions. Somehow we must weaken the constraints placed upon the function by differentiability. 

In the many guises it is used, \cref{fund} captures a fundamental relation of calculus, but is only known to hold in rather special situations. 
If $f,g,z\in\reals$, or if $f,g,z\in\coms$ and $g$ is analytic, then \cref{fund} holds by the {\it fundamental theorem of calculus}. 
If $g$ and $z$ are vectors then the products and derivatives must be of restricted type, for instance if $f\in\reals$ with $g=\nabla f$, then \cref{fund} is just known as the `gradient theorem'. 
Differential geometry provides certain other instances of \cref{fund} provided by inner or outer products between $dz$ and $g$. 
We will define the direct algebraic product via the {\it mixture}.

Of course, there is good reason why the derivative $\sfrac{df}{dz}$ in general belongs to a larger space than $z$ and $f$. For multi-dimensional objects $f$ and $z$, the derivative has many roles to play as the directional derivative, divergence, curl, the Jacobian, or the Lie derivative. In no interpretation can these be packaged up into a single object $df/dz$ with the same dimensionality as $z$ and $f$. There is moreover no obvious geometrical way to derive a rate of change $df/dz$ as a limiting quantity, with a unique value, if $z$ and $f$ are multi-dimensional, such that $z,f,f/z,df/dz\in\op S$. 

Rather than precluding the existence of such a derivative, this suggests that the analytic closure we seek (of $z,f,g$ all belonging to the same space) imposes strong restrictions on a function, too strong admit all but trivial functions in general, unless we find them extra freedom by letting their underlying space curve. 

If we allow curvature of the underlying coordinate system, we obtain the freedom needed for \cref{Taymin} to admit non-trivial functions. As we move around in $z$-space, the basis itself varies in such a way as to compensate for any variation of $f$, and ensure that the equations \cref{Taymin} remain satisfied. 

Thus in seeking a coordinate system in which $f$ can be expressed analytically, $f$ itself becomes a `source' term for curvature of the underlying space. If we seek such functions in describing the physical world, then as a consequence we perceive that world as curved by a system of forces `tensioned' by seeking functions that are differentiable. 

The problems of extending calculus into higher dimensions are well known. Most subtle perhaps is that to describe the change in a function, $df(x_0)=f(x)-f(x_0)$, is problemmatic because we must understand how the space of $f$ at $x$ is related to the space of $f$ at $x_0$, which is non-trivial if the underlying space can curve, and so {\it tensor} calculus accounts for the variation of the basis from one point to another, i.e. from $x$ to $x_0$. We may attribute this variation to an external source --- the mass-energy terms of general relativity for instance --- but here we will fix the source of variation as just that curvature necessary to achieve a strong form of differentiability. This `strong' form says that a function $f$ of a multi-dimensional variable $x$ is differentiable with respect to $x$ as a whole, and not merely partially differentiable with respect to its components, similar to the concept of analyticity of complex functions. We therefore refer to the resulting variation as {\it analytic curvature}.

\subsection{Differentiability and analyticity}

Let us seek the kind of functions $f(z)$ for which $z$, $f$, and $\sfrac{df}{dz}$ may belong to the same algebraic space, according to a series expansion of $f$. That is, given a set $\op S$ spanned by bases $\e0,\e1,\dots,\e n$, we require closure with respect to a function $f$ and variable $z$, wherein given $z\in\op S$ and $f\in\op S$, the algebraic operations $z\pm f$, $zf$, and $z/f$, lie in $\op S$, and there exists a unique derivative $df/dz\in\op S$ corresponding to the limiting ratio of infinitesimals $\delta f\in\op S$ and $\delta z\in\op S$. 

Suppose there exists an operator $\dagger$ with which 
the series expansion of $f$ can be written as
\begin{equation}\label{diff1}
f=\left.f\right|_{\delta z=0}+\bb{\delta z\sfrac{\partial\;}{\partial z}+\delta z^\dagger \sfrac{\partial\;}{\partial z^\dagger }}\left.f\right|_{\delta z=0}+\ord{|\delta z|^2}\;.
\end{equation}
%
If a coordinate system can then be found in which 
\begin{equation}\label{cr1}
\sfrac{\partial\;}{\partial z^\dagger }f=0\;,
\end{equation}
then \cref{diff1} reduces to
\begin{equation}\label{diff0}
f=\left.f\right|_{\delta z=0}+\delta z\sfrac{\partial\;}{\partial z}\left.f\right|_{\delta z=0}+\ord{|\delta z|^2}\;.\qquad\qquad\quad
\end{equation}
After multiplying \cref{diff0} from the left by the inverse $\delta z^{-1}$, we can now define the derivative of $f$ with respect to $z$, if \cref{cr1} holds, as
\begin{align}\label{dfdz}
f':=\lim_{|\delta z|\rightarrow0}\delta z^{-1}(f-\left.f\right|_{\delta z=0})
=\sfrac{\partial\;}{\partial z}\left.f\right|_{\delta z=0}\;,\qquad\quad
\end{align}
with higher order terms of \cref{diff0} vanishing since $$\displaystyle\lim_{|\delta z|\rightarrow0}\delta z^{-1}\ord{|\delta z|^2}=\lim_{|\delta z|\rightarrow0}\ord{|\delta z|}=0\;.$$
The condition \cref{cr1} will thus provide a function satisfying \cref{Taymin}. 

We must, therefore, first understand how to expand a function in the form \cref{diff1}. 
If $z,f\in\coms$ and the operator $\dagger$ is the complex conjugate, then \cref{cr1} is the set of Cauchy-Riemann equations. 
In higher dimensions we must further decompose the derivative to obtain \cref{diff1}, as follows. 

Assuming $\delta z\sfrac{\partial\;}{\partial z}$ has both vector and scalar parts, we extract the scalar part by adding the mirror of $\delta z\sfrac{\partial\;}{\partial z}$. Applying this as an operator to $f$,
\begin{align}
\bb{\delta z\sfrac{\partial\;}{\partial z}+\mirror{\delta z\sfrac{\partial\;}{\partial z}}}f
&=\delta z\bb{\sfrac{\partial\;}{\partial z}+\delta z^{-1}\mirror{\sfrac{\partial\;}{\partial z}}\mirror{\delta z}}f\nonumber\\
:\!&=\;\delta z\bb{\sfrac{\partial\;}{\partial z}f+\con(f)}\label{dzpartial}
\end{align}
defining
\begin{align}
\con(f)&=\delta z^{-1}\mirror{\sfrac{\partial\;}{\partial z}}\mirror{\delta z}\;f\;.\label{conf}
\end{align}
We define the quantity multiplying $\delta z$ on the righthand-side of \cref{dzpartial} as a proper derivative, denoted 
\begin{align}\label{properder}
\sfrac{d\;}{dz}f=\sfrac{\partial\;}{\partial z}f+\con(f)\;.
\end{align}

This derivative is not unique because the quantity $\con(f)$ depends on the direction in which $\delta z\rightarrow0$. We will identify these different possible values of $\con(f)$ with different coordinate systems, by identifying $\sfrac{d}{dz}$ with the covariant differential operator.

The differentiability condition \cref{cr1}, in covariant terms, becomes
\begin{align}
0
&=d\adj z\sfrac{d\;}{d\adj z}f\nonumber\\
&=\e{\adj\gamma} dz^{\adj\gamma}\ec{\adj\beta}\e\alpha f^\alpha_{;\adj\beta}\nonumber\\
&=\e{\adj\gamma}\delta z^{\adj\gamma}\e\mu\eta^{\adj\beta\mu}_\alpha\bb{\sfrac{\partial\;}{\partial z^{\adj\beta}}f^\alpha+\con^\alpha_{\lambda\adj\beta}f^\lambda}
\end{align}
which holds for any $d\adj z=\e{\adj\gamma} dz^{\adj\gamma}$ if $
0
=\e\mu\eta^{\adj\beta\mu}_\alpha f^\alpha_{;\adj\beta}\nonumber
$, and since this must vanish for each $\mu$-indexed component we have
\begin{align}\label{anacond}
0=\eta^{\adj\beta\mu}_\alpha f^\alpha_{;\adj\beta}
\;.
\end{align}
We call this the {\it analyticity condition}. 

In essence \cref{anacond} is the extension of the Cauchy-Riemann equations to our covariant geometries, and indeed it reduces to them for complex functions. 

There are now 3 prongs to this calculus: the algebra defined by the mixture $\eta$, curvature of that algebra defined by the connection $\con$, and the class of analytic functions $f$ so permitted.

The source terms for the curvature measured by $\con(f)$ are the mirror derivatives of $f$. To find these write
\begin{align}
\sfrac{d\;}{dz}=\sfrac1{n}\ec\alpha\sfrac{\partial\;}{\partial z^\alpha}\;,
\end{align}
on an $n$-dimensional space. We then have
\begin{align}
n\sfrac{d}{dz}f=\ec\beta\sfrac{d\;}{dz^\beta}\e\alpha f^\alpha
=\ec\beta\e\alpha\bb{\sfrac{\partial\;}{\partial z^\beta}f^\alpha+\con^\alpha_{\lambda\beta}f^\lambda}\;.
\end{align}
If we let $r=\delta z/|\delta z|$, then $\delta z^{-1}=\mirror{\delta z}/|\delta z|=\mirror r$, giving 
\begin{align}
\ec\beta\e\alpha\con^\alpha_{\lambda\beta}f^\lambda
&=\e{\mirror\nu} \ec{\mirror\beta}\e{\mirror\gamma}\e\alpha r^\nu r^\gamma f^\alpha_{,\beta}
\end{align}
or in terms of the mixture, 
\begin{align}
\e\gamma\eta^{\gamma\beta}_\alpha\con^\alpha_{\lambda\beta}f^\lambda
&=\e\gamma\eta^\gamma_{\mirror\nu\kappa}\eta^{\kappa\mirror\beta}_\delta\eta_{\mirror\lambda\alpha}^\delta r^\nu r^\gamma f^\alpha_{,\beta}\nonumber\\
&=\e\gamma\eta^\gamma_{\mirror\nu\kappa}\eta^{\kappa\mirror\beta}_\delta\eta_{\mirror\sigma\alpha}^\delta r^\nu r^\gamma \sfrac1{|f|}f_\lambda f^\alpha_{,\beta} f^\lambda\;.
\end{align}
Omitting the basis $\e\gamma$, this is just a matrix-vector equation, 
\begin{align*}
[..lhs..]^\gamma_\lambda f^\lambda
=[..rhs..]^\gamma_\lambda f^\lambda
\quad\Rightarrow\quad
[..lhs..]^\gamma_\lambda
=[..rhs..]^\gamma_\lambda\;.
\end{align*}
Let us assume we can write $\con^\alpha_{\lambda\beta}=\sfrac1{ n}\eta^\alpha_{\beta\omega}h_\lambda^\omega$ for some sourve vector $h$, and assume the mixture has then inverse \cref{etainverse}, then the `$lhs$' term becomes $\eta^{\gamma\beta}_\alpha\con^\alpha_{\lambda\beta}=\eta^{\gamma\beta}_\alpha(\sfrac1{\im n}\eta^\alpha_{\beta\omega}h_\lambda^\omega)=h_\lambda^\gamma$, so we have
\begin{align}
h_\lambda^\gamma
&=\eta^\gamma_{\mirror\nu\kappa}\eta^{\kappa\mirror\beta}_\delta\eta_{\mirror\sigma\alpha}^\delta r^\nu r^\gamma \sfrac1{2|f|}f_\lambda f^\alpha_{,\beta}\;.
\end{align}

\subsection{Integrability: (overtly) illustrative examples}

Let us take a somewhat artificial but accessible example illustrating the above. 

Take bases $\e1$ and $\e2$, and a mixture product such that $\e1^2=\e2^2=1$ and $\e1\e2=-\e2\e1$. 
Consider the simple function $g(x,y)=x^2+y^2$, and integrate from $(x,y)=(0,-1)$ to $(0,+1)$ with respect to a vector variable $\v r=x\e1+y\e2$, 
\begin{align}\label{pareg}
\int_{-\e2}^{+\e2}\!\!\!\!\!\! d\v r|\v r|^2=\int_{(0,-1)}^{(0,+1)}\!\!\!\!\!\! (dx\e1+dy\e2)(x^2+y^2)\;,\qquad\qquad
\end{align}
where $\e1$ and $\e2$ denote the coordinate bases. (Note that we put the `$d\v r$' first in the integral for consistency later). The integrand is a regular function, 
yet the integral has no unique solution. 
For example if we integrate along a piecewise linear path, anti-clockwise around a rectangle with sides $(0,\pm1)$ and $(c,\pm1)$, the horizontal segments cancel each other out leaving
\begin{align}
\int_{-\e2}^{+\e2}\!\!\!\!\!\! d\v r|\v r|^2&
=\int_{-1}^{+1}dy\e2(c^2+y^2)
\;=\;2(c^2+\sfrac13)\e2\;,\quad\;
\end{align}
which gives a different result for every value of $c$. If instead we integrate anti-clockwise around a semicircular arc between $(0,\pm1)$, we have 
\begin{align}
\int_{-\e2}^{+\e2}\!\!\!\!\!\! d\v r|\v r|^2&
=\int_{-\pi/2}^{+\pi/2}(-\sin\theta\e1+\cos\theta\e2)d\theta
\;=\;2\e2\;.
\end{align}
Although we can write $g=\sfrac{\partial\;}{\partial\v r}\cdot\v f$ for some $\v f$, the integrand $d\v rg=d\v r\sfrac{\partial\;}{\partial\v r}\cdot\v f$ does not then equal $d\v f$, and the integral is not unique. In some sense this is because we are missing information. 
If instead we have an algebraic product with which we can instead write $g=\sfrac{\partial\;}{\partial\v r}\v f$ (without the dot product between the derivative and $\v f$), and moreover we can define a function $\con(\v f)=d\v r^{-1}\sfrac{\partial\;}{\partial\v r}d\v r\;\v f$, then the augmented integral gives
\begin{align}\label{gright}
\int_a^b d\v r \big(g+\con(f)\big)&=\int_a^b (d\v r \sfrac{\partial\;}{\partial\v r}+\sfrac{\partial\;}{\partial\v r}d\v r)\v f\;,\nonumber\\
&=\int_{\v f(a)}^{\v f(b)}d\v f\;=\;\v f(b)-\v f(a)\;,
\end{align}
in agreement with \cref{fund}. That is, the integrand of \cref{gright} {\it does} equal $d\v f$ and yields path independence. 
The first part of the integral on the lefthand side of \cref{gright} evaluates as
\begin{subequations}
\begin{align}
\int_a^b d\v r \sfrac{\partial\;}{\partial\v r}\v f&
=\int_{(0,-1)}^{(0,+1)}(dx\e1+dy\e2)(\sfrac{\partial\;}{\partial x}\e1+\sfrac{\partial\;}{\partial y}\e2)\sfrac13(x^3\e1+y^3\e2)\nonumber\\&
=\int_{(0,-1)}^{(0,+1)}(dx\e1+dy\e2)(x^2+y^2)\;,
\end{align}
while the second part, $\int_a^bd\v r \con(\v f)$, evaluates as
\begin{align}
\int_a^b \sfrac{\partial\;}{\partial\v r}d\v r\hspace{0.05cm}\v f&
=\int_{(0,-1)}^{(0,+1)}(\sfrac{\partial\;}{\partial x}\e1+\sfrac{\partial\;}{\partial y}\e2)(dx\e1+dy\e2)\sfrac13(x^3\e1+y^3\e2)\nonumber\\&
=\int_{(0,-1)}^{(0,+1)}(dx\e1-dy\e2)(x^2-y^2)\;.
\end{align}
\end{subequations}
Their sum (which we can calculate before or after integrating them) is 
\begin{align}
\int_a^b (d\v r \sfrac{\partial\;}{\partial\v r}+\sfrac{\partial\;}{\partial\v r}d\v r)\v f&
=2\int_{(0,-1)}^{(0,+1)}dx\e1x^2+dy\e2y^2\nonumber\\&
=\begin{cases}
2\int_{-1}^{+1}dy\e2y^2&(i)\\
\int_{-\pi/2}^{+\pi/2}(\e2\sin\theta-\e1\cos\theta)\sin2\theta d\theta&(ii)
\end{cases}\nonumber\\&
=\sfrac43\e2
\end{align}
where we integrate around the rectangle in (i) and the semicircular arc in (ii), obtaining the unique result of $\sfrac43\e2$.

That is all very well, but in what sense have we solved the original problem? We have said that $g=|\v r|^2$ can be considered the derivative of the function $\v f$ if we allow for the by-product $\con(\v f)=d\v r^{-1}\sfrac{\partial\;}{\partial\v r}d\v r\v f$ in that differentiation. Since that by-product is not directly associated with the function $g$, we say that is associated with the underlying space instead. 

A similar procedure can be used for complex functions. Taking the same function $g=x^2+y^2$ as above, replacing $\e1$ with unity and $\e2$ with $\im=\sqrt{-1}$, the results above all follow similarly, but in addition with $\sfrac{\partial\;}{\partial\conj z}\v f=-\conj\con(\v f)=x^2-y^2$, we find that $\v f$ satisfies the Cauchy-Riemann equations in the form $\big(\sfrac{\partial\;}{\partial\conj z}\v f+\conj\con(\v f)\big)=0$ with the $\con$ by-product of differentiation included. 

We can also absorb the residues of complex loop integrals of meromorphic functions into a by-product $\con$, for example when integrating $g=1/z$ over an anti-clockwise circle $C$ centred on $z=0$ in the complex plane. In that case the two integrals we must consider are 
\begin{align}
\oint_Cdz/z&=2\pi\im\qquad{\rm and}\qquad\oint_Cd\conj z/\conj z=-2\pi\im\;.
\end{align}
In this case the integrand on the left can be written as $1/z=\sfrac{\partial\;}{\partial z}f$ where $f=2\log|z|$, while the integrand on the right we define as $dz\con(f)$, where $\con(f)=dz^{-1}d\conj z/z^*$ is again interpreted as a by-product of differentiating $f$. The sum of the two integrals then vanishes in accordance with the fundamental theorem. 

These are, as I said above, overt examples with a somewhat artificial error or `curvature' term that is easy to deconstruct. In higher dimensions we face the problem that {\it no} functions are differentiable in the sense we seek, and hence no functions are integrable, unless we permit the existence of the kind of by-products represented by $\con$. More important we are faced, in applications, with seeking empirical fields with which to study the physical world, whose ideal forms are a priori unknown, and therefore any by-products that would be required are less obvious, and less artificial than those above. When the by-product is not any obvious part of the function we are studying, we may instead assume that it is created by the underlying space, and ask where such a product might come from. This is the proposal of {\it analytic curvature}.

\section{Physical laws}\label{sec:physics}

The algebra and calculus set out above provide notably more freedom than the usual approach to covariant calculus, because of the role played by the mixture. They place gauge freedom at the centre of everything, so it should be no surprise that various fundamental laws find elegant expression, but we shall go a little further, exploring the extent to which physical laws are all just expressions of analyticity of the functions we choose to represent physical quantities. 

\subsection{Curvature from simple fields: electromagnetic \& beyond}\label{sec:standard}

Let us look at the basic forms of differential expressions that arise from variation of the basis using mixture algebra. 

The derivative of a basis is given as per \cref{edcon} by $\e{\alpha,\beta}=\con^\gamma_{\alpha\beta}\e\gamma$ and $\ec\alpha{,\beta}=\con^\alpha_{\gamma\beta}\ec\gamma$. We will return to this shortly, but first consider a simplification in which $\e{\alpha,\beta}=h_\beta\e\alpha$ and $\ec\alpha_{,\beta}=h_\beta\ec\alpha$. From this we can find the connection, 
\begin{align}\label{hcon}
\e{\alpha,\beta}=h_\beta\e\alpha  \quad
\Rightarrow \quad \e\mu\con^\mu_{\alpha\beta}=\e\mu \id^\mu_\alpha h_\beta \quad
\Rightarrow \quad \con^\mu_{\alpha\beta}=\id^\mu_\alpha h_\beta\;.
\end{align}
Note in particular that the divergence of the basis is
\begin{align}\label{hdiv}
\ec\alpha_{,\alpha}=h_\alpha\ec\alpha\qquad {\rm or}\qquad g^{\alpha\beta}\e{\alpha,\beta}=h^\alpha\e\alpha\;.
\end{align}
Let this be identified with the electromagnetic potential $h_\alpha=\cc{\phi,\v A}$, and note this is then just the trace of the connection, 
\begin{align}
h_\alpha=\con^\beta_{\alpha\beta}\;.
\end{align} 
Let us then look at some general expressions characterizing its calculus. 

%
%
To find the curvature induced by this source, take the second derivative of the basis, 
\begin{align}
\e{\alpha,\mu\nu}&=(h_\mu\e\alpha)_{,\nu}\nonumber
=\id^\gamma_\alpha(h_{\mu,\nu}+h_\mu h_\nu)\e\gamma
\end{align}
wihch is equal to $\scon^\gamma_{\alpha\mu\nu}\e\gamma$ by \cref{fR}, hence
\begin{align}
P^\gamma_{\alpha\mu\nu}&=\id^\gamma_\alpha(h_{\mu,\nu}+h_\mu h_\nu)\quad
\Rightarrow\quad R^\gamma_{\alpha\mu\nu}=\id^\gamma_\alpha h_{[\mu,\nu]}=\id^\gamma_\alpha F_{\nu\mu}\;,
\end{align}
defining a field tensor $F_{\mu\nu}$ consistent with the standard electromagnetic field (Faraday) tensor. This result is reminiscent of Pauli's association of the Riemann tensor $R$ with the Faraday tensor $F$ (see e.g. \cite{dawn}). 
%
%
The derivative $f'=\ec\beta\partial_\beta(f^\alpha\e\alpha)=\e\gamma\eta^{\gamma\beta}_\alpha f^\alpha_{;\beta}$ then has components
\begin{align}
\eta^{\gamma\beta}_\alpha f^\alpha_{;\beta}\nonumber
&=\eta^{\gamma\beta}_\alpha(f^\alpha_{,\beta}+f^\lambda\con^\alpha_{\lambda\beta})\nonumber\\
&=\eta^{\gamma\beta}_\alpha(f^\alpha_{,\beta}+f^\lambda1_\lambda^\alpha h_{\beta})\nonumber\\
&=\eta^{\gamma\beta}_\alpha(\partial_{\beta}+h_{\beta})f^\alpha\;.
\end{align} 

In the simplified case \cref{hcon} the effect of the derivative $\partial_\beta$ on the basis is to multiply it by the potential $h_\beta$. 
In the most general case is given by \cref{fR}, where instead 
the derivative has components
\begin{align}\label{diracstand}
\eta^{\gamma\beta}_\alpha f^\alpha_{;\beta}
&=\eta^{\gamma\beta}_\alpha(f^\alpha_{,\beta}+f^\lambda\con^\alpha_{\lambda\beta})\nonumber\\
&=(\eta^{\gamma\beta}_\alpha \partial_{\beta}+\eta^{\gamma\beta}_\lambda\con^\lambda_{\alpha\beta})f^\alpha\nonumber\\
&=(\eta^{\gamma\beta}_\alpha \partial_{\beta}+ H^\gamma_\alpha)f^\alpha\;,
\end{align}
for some $H^\gamma_\alpha$, which we will return to in \cref{sec:dirac}, but shall see is consistent with the Dirac equation. 
If $\con$ is symmetric then let us re-write this as
\begin{align}\label{dstand}
\eta^{\gamma\beta}_\alpha f^\alpha_{;\beta}
&=\eta^{\gamma\beta}_\alpha(f^\alpha_{,\beta}+f^\lambda\con^\alpha_{\beta\lambda})\nonumber\\
&=(\eta^{\gamma\beta}_\alpha \partial_{\beta}+\eta^{\gamma\beta}_\lambda\con^\lambda_{\beta\alpha})f^\alpha\nonumber\\
&=\eta^{\gamma\beta}_\lambda(1^\lambda_\alpha\partial_{\beta}+(\op G_{\beta})^\lambda_\alpha)f^\alpha\nonumber\\
&=\eta^{\gamma\beta}_\lambda(1\partial_{\beta}+\op G_{\beta})^\lambda_\alpha f^\alpha\;,
\end{align}
where $\op G_\mu$ is a square matrix with components $(\op G_\mu)^\alpha_\beta=\con^\alpha_{\mu\beta}$. This is consistent with the derivative of the standard model of particle physics (see e.g. \cite{cg07,yangmills}), letting $\op G_\mu=\epsilon\op H_\mu$ where $\op H_\mu$ are field potentials and $\epsilon$ a coupling parameter. These matrix expressions are clearly capable of exhibiting the symmetries of the SU(3)$\times$SU(2)$\times$U(1) group. 

The curvature tensor is then
\begin{align}
R^{\alpha}_{\beta\nu\mu}&=\con^{\alpha}_{\mu\beta,\nu}-\con^{\alpha}_{\nu\beta,\mu}+\con^{\alpha}_{\nu\sigma}\con^{\sigma}_{\mu\beta}-\con^{\alpha}_{\mu\sigma}\con^{\sigma}_{\nu\beta}\nonumber\\
&=\epsilon(\op H^{}_{\mu,\nu}-\op H^{}_{\nu,\mu}+\epsilon\op H^{}_{\nu}\op H^{}_{\mu}-\epsilon\op H^{}_{\mu}\op H^{}_{\nu})^\alpha_\beta\nonumber\\
&=2\epsilon(\op H^{}_{[\mu,\nu]}+\epsilon\op H^{}_{[\nu}\op H^{}_{\mu]})^\alpha_\beta
\;:=\;2\epsilon(F_{\mu\nu})^\alpha_\beta\;,
\end{align}
which seems consistent with the field tensor $F^{\mu\nu}$ of the Yang-Mills theory \cite{yangmills}, if we assume potentials $\op H_\mu$ consistent with the SU(3) group (whereas in electromagnetic theory the field tensor is simply $F_{\mu\nu}=\op H_{[\mu,\nu]}$). 
%
%
In particular we can show that this satisfies the transformation properties of the Yang-Mills theory. Following the argument in \cite{yangmills}, consider a wave-function $f$, transformed under an isotopic gauge transformation $S$ as
\begin{align}
f&=Sf'
\end{align}
with invariance implying
\begin{align}
S(\partial_\mu-\im\eps \op H_\mu')f'&=(\partial_\mu-\im\eps \op H_\mu)f\;,
\end{align}
from which two conditions we can show that 
\begin{align}
\op H_\mu'&=S^{-1}\op H_\mu S+\sfrac\im\eps S^{-1}\partial_\mu S\;.
\end{align}
We can then show that the field tensor as derived above transforms correctly as
\begin{align}
F_{\mu\nu}'=S^{-1}F_{\mu\nu}S\;.
\end{align}
This takes some lengthy but straightforward algebra, but we indeed have
\begin{align}
F_{\mu\nu}'&=\op H_{[\mu,\nu]}'+\im\eps \op H_{[\mu}' \op H_{\nu]}'\nonumber\\
		 &=S^{-1}\bb{\op H_{[\mu,\nu]}+\im\eps \op H_{[\mu} \op H_{\nu]}}S\nonumber\\
		 &=S^{-1}F_{\mu\nu}S\;.
\end{align}

Thus we see certain elements of a rich calculus that appears to have application to, for example, the standard model of particle physics, arising from the derivatives and curvatures expressed in this fashion, requiring little more than standard covariant calculus, but crucially involving the mixture (as well as the connection), from which spring the various possible symmetry groups in which these relations may be expressed. 
Of course these are outline expressions, and we have not seriously turned them to such applications in any depth, but these may suggest some worth in exploring the properties of the fields implied by these, in particular their masses and interactions, the possible symmetries of course, and extending these ideas to obtain variational expressions such as Lagrangians. These lie beyond my ambitions here, but we should at least probe whether the aesthetic similarity of the expressions above indeed correspond to the known physical relations. So let us take \cref{diracstand} and show that this derivative indeed produces the Dirac equation of a spin $\hf$ particle.

\subsection{The covariant Dirac equation}\label{sec:dirac}

The covariant derivative of a function $f$ is 
\begin{align}\label{fder}
f'=\sfrac{\partial\;}{\partial z}f
&= \e\gamma\eta_{\alpha}^{\gamma\beta} f^\alpha_{;\beta}\nonumber\\
&= \e\gamma\eta_{\alpha}^{\gamma\beta} (f^\alpha_{,\beta}+f^\lambda\con_{\lambda\beta}^\alpha)\nonumber\\
&= \e\gamma(\eta_{\alpha}^{\gamma\beta} \partial_{\beta}+\eta_{\lambda}^{\gamma\beta} \con_{\alpha\beta}^\lambda)f^\alpha\;.
\end{align}
Here we can factorize out either the mixture (second line) or the components of $f$ (third line). 
Taking the latter, let $\eta_{\lambda}^{\gamma\beta} \con_{\alpha\beta}^\lambda=\n H^\gamma_\alpha$ for some scalar $\n$ and some normalized tensor $ H^\gamma_\alpha$, then
\begin{align}\label{fdern}
f'&= \e\gamma(\eta_{\alpha}^{\gamma\beta} \partial_{\beta}+\n H^\gamma_\alpha)f^\alpha\;.
\end{align}
Furthermore let $\eta_{\lambda}^{\gamma\adj\beta} \con_{\alpha\adj\beta}^\lambda=\n\hat H^\gamma_\alpha$ for some $\hat H$, so overall
\begin{subequations}\label{diraccon}
\begin{align}
\eta_\alpha^{\gamma\beta}\con_{\mu\beta}^\alpha&=\n H^\gamma_\mu\qquad\Rightarrow&\hspace{-0.9cm}
\con_{\mu\beta}^\alpha&=\sfrac1{ n}\n\eta^\alpha_{\beta\gamma} H^\gamma_\mu\;,\\
\eta_\alpha^{\gamma\adj\beta}\con_{\mu\adj\beta}^\alpha&=\n\hat H^\gamma_\mu\qquad\Rightarrow&\hspace{-0.9cm}
\con_{\mu\adj\beta}^\alpha&=\sfrac1{ n}\n\eta^\alpha_{\adj\beta\gamma}\hat H^\gamma_\mu\;.
\end{align}
\end{subequations}


If the adjoint derivative of $f$ vanishes by \cref{anacond}, then we have
\begin{align}\label{diraceq}
0= \sfrac{\partial\;}{\partial\adj z}f
&=\eta_{\alpha}^{\gamma\adj\beta} f^\alpha_{;\adj\beta}\nonumber\\
&= \eta_{\alpha}^{\gamma\adj\beta} (f^\alpha_{,\adj\beta}+f^\lambda\con_{\lambda\adj\beta}^\alpha)\nonumber\\
&= (\eta_{\alpha}^{\gamma\adj\beta} \partial_{,\beta}+\n\hat H^\gamma_\alpha)f^\alpha\;.
\end{align}
This looks superficially like Dirac's equation for an electron of mass $\m$ if $\n=\m c/\im\hbar$, but let us probe further. 

If the adjoint derivative of $f'$ in \cref{fder} also vanishes, expanding out time-like and space-like ($0$ and $1,2,3$ index) parts of the derivative $\sfrac{\partial\;}{\partial\adj z}f'$ gives
\begin{align}
0
&=\e\gamma(\eta_\alpha^{\gamma\adj\beta}\partial_{\adj\beta}+\eta_\lambda^{\gamma\adj\beta}\con_{\alpha\adj\beta}^\lambda)(\eta^{\alpha\omega}_\sigma\partial_{\omega}+\eta^{\alpha\omega}_\nu\con^\nu_{\sigma\omega})f^\sigma\nonumber\\
&=\e\gamma(\eta^{\gamma\adj\beta}_\alpha\partial_{\adj\beta}+\n \hat H^\gamma_\alpha)(\eta^{\alpha\omega}_\sigma\partial_{\omega}+\n{ H}^\alpha_\sigma)f^\sigma\qquad\qquad\qquad\quad\mbox{[by \cref{diraccon}]}\nonumber\\
&=\e\gamma(\eta^{\gamma0}_\alpha\partial_{0}-\eta^{\gamma i}_\alpha\partial_{i}+\n \hat H^\gamma_\alpha)(\eta^{\alpha0}_\sigma\partial_0+\eta^{\alpha j}_\sigma\partial_j+\n H^\alpha_\sigma)f^\sigma\quad\hspace{0.08cm}\mbox{[by \cref{def:adjoint}]}\nonumber\\
&=\e\gamma\left(\;\;
\eta^{\gamma0}_\alpha\eta^{0\alpha}_\sigma\partial_0^2
-\eta^{\gamma i}_\alpha\eta^{\alpha j}_\sigma\partial_{i}\partial_j
\;+\;\n^2 \hat H^\gamma_\alpha H^\alpha_\sigma\right.\nonumber\\
&\qquad
+\;\cc{\eta^{\gamma0}_\alpha\eta^{\alpha i}_\sigma-\eta^{\gamma i}_\alpha\eta^{\alpha0}_\sigma}\partial_{0}\partial_i\;+\;\eta^{\gamma\adj\beta}_\alpha\cc{\eta^{\alpha\omega}_{\sigma,\adj\beta}\partial_\omega+\n{ H}^\alpha_{\sigma,\adj\beta}}\\
&\left.\qquad+\;\;\n\!\{\hat H^\gamma_\alpha\eta^{\alpha0}_\sigma+\eta^{\gamma0}_\alpha H^\alpha_\sigma\}\partial_{0}
\;+\;\n\!\{\hat H^\gamma_\alpha\eta^{\alpha i}_\sigma-\eta^{\gamma i}_\alpha H^\alpha_\sigma\}\partial_i
\;\;\;\;\right)f^\sigma\;.\nonumber
\end{align}

If the last two lines of this equation vanish then we obtain the Klein-Gordan equations with a source term, and then the analyticity condition \cref{diraceq} is indeed the Dirac equation. This implies the algebraic conditions
\begin{align}\label{KGcond}
\eta^{\gamma0}_\alpha\eta^{\alpha0}_\sigma=\eta^{\gamma i}_\alpha\eta^{\alpha i}_\sigma&=1^\gamma_\sigma&
H^\gamma_\alpha\hat H^\alpha_\sigma&=-1^\gamma_\sigma\nonumber\\
\eta^{\gamma0}_\alpha\eta^{\alpha i}_\sigma-\eta^{\gamma i}_\alpha\eta^{\alpha0}_\sigma&=0&
\hat H^\gamma_\alpha\eta^{\alpha i}_\sigma-\eta^{\gamma i}_\alpha H^\alpha_\sigma&=0\\
\hat H^\gamma_\alpha\eta^{\alpha0}_\sigma+\eta^{\gamma0}_\alpha H^\alpha_\sigma&=0&
\left.\eta^{\gamma i}_\alpha\eta^{\alpha j}_\sigma+\eta^{\gamma j}_\alpha\eta^{\alpha i}_\sigma\right|_{i\neq j}&=0\nonumber
\end{align}
plus a derivative condition
\begin{equation}\label{dirdiv}
\eta^{\gamma\beta}_\alpha\cc{\eta^{\alpha\omega}_{\sigma,\beta}\partial_\omega+\n{\hat H}^\alpha_{\sigma,\beta}}f^\sigma=0\;.
\end{equation}
The latter is satisfied trivially if $\eta^{\alpha\omega}_{\sigma,\beta}={\hat H}^\alpha_{\sigma,\beta}=0$. 


If $\eta^{\gamma0}_\alpha$ is the identity matrix $1^\gamma_\alpha$ then the penultimate condition of \cref{KGcond} implies $\hat H=-H$, and then the matrices $\eta^{\gamma1}_\alpha$, $\eta^{\gamma2}_\alpha$, $\eta^{\gamma3}_\alpha$, $H^\gamma_\alpha$, behave algebraically as the Dirac matrices, writing the conditions above as matrix equations
\begin{align}\label{KGcondmat}
\tens1&=\tens\eta^0\tens\eta^0=\tens\eta^i\tens\eta^i=
\tens H\hspace{0.02cm}{\tens H}\;,\nonumber\\
\tens0&=\tens\eta^0\tens\eta^i-\tens\eta^i\tens\eta^0=
\tens H\hspace{0.02cm}\tens\eta^0-\tens\eta^0{\tens H}\;,\\
\tens0&=\left.\tens\eta^i\tens\eta^j+\tens\eta^j\tens\eta^i\right|_{i\neq j}=
\tens\eta^i{\tens H}+\tens H\hspace{0.02cm}\tens\eta^i\;.\nonumber
\end{align}
For example we may associate the mixture with `block' or tensor products of the Pauli matrices $\sigma_\alpha$ as $\eta^{\gamma0}_\alpha=\id^\gamma_\alpha$ and
$\eta^{\gamma1}_\alpha=\sq{\sigma_1\otimes\sigma_2}^\gamma_\alpha$\;, 
$\eta^{\gamma2}_\alpha=\sq{\sigma_2\otimes\sigma_0}^\gamma_\alpha$\;, 
$\eta^{\gamma3}_\alpha=\sq{\sigma_1\otimes\sigma_1}^\gamma_\alpha$\;, 
$H^\gamma_\alpha=\sq{\sigma_1\otimes\sigma_3}^\gamma_\alpha$\;. 

Thus writing $N=Mc/\im\hbar$ we have Dirac's equations \cite{dirac28,diracqm} for a spin $\hf$ particle of mass $\m$, 
\begin{align}
0=\bb{\eta_\alpha^{\gamma\adj\beta}\partial_{\adj\beta}-\sfrac{\m c}{\im\hbar} H^\gamma_\alpha}f^\alpha\quad\;
\end{align}
along with the corresponding Klein-Gordan equation, 
\begin{align}
0&
=\bb{{\partial_0^2-\partial_i^2}+\sfrac{\m^2c^2}{\hbar^2}}f^\alpha\;.
\end{align}

The mixture tensors $\eta$ implied by this define a set of bases $\e\alpha$ or $\ec\alpha$ which, using the relation $\ec\gamma\ec\beta=\eta^{\gamma\beta}_\alpha\ec\alpha$, are non-associative. 
This hints at a very different geometry at work for a system that satisfies the Dirac and Klein-Gordon equations, compared to something like the natural geometry of familiar macroscopic space-time. That geometry results from the strong restriction of satisfying the Klein-Gordon equations, which may suggest they can be expected only to hold weakly in some sense, for example locally, i.e. on small scales. 
\subsection{Quantization and field potentials}

%
%

Following on from Dirac's equations as derived above, note that if two solutions $f$ and $\tilde f$ have an offset $\phi$ in their phase, 
\begin{equation}
f= \tilde f\exp{-\im\phi}
\end{equation}
their derivatives are related by
\begin{equation}
\sfrac{\partial\;}{\partial z^\alpha}f= \bb{\sfrac{\partial\;}{\partial z^\alpha}+\im\sfrac{\partial\phi}{\partial z^\alpha}}\tilde f\;.
\end{equation}
Let us say $\phi=z^\beta w_\beta/\hbar$. If we take a loop around the origin of the $(z^0,z^i)$ system, the phase must equal a multiple of $2\pi$, 
\begin{equation}\label{diracint}
\oint dz z^\alpha(z+\omega)_\alpha=\oint dz z^\alpha\omega_\alpha=2\pi nRes[z\cdot w,z=0]:=e(\phi+A)
\end{equation}
with $e$ being a multiple of $n\in\mathbb Z$ (e.g. letting $e=n\lambda$ with $\lambda\in\mathbb R$ and $\phi+A=\sfrac{2\pi}\lambda Res[z\cdot w,z=0]\in\mathbb R$). Thus, as Dirac showed \cite{dirac1931}, the momenta are unique only up to a quantized shift. If we let $\tilde f\sim \exp{-(z^0z_0+z^iz_i)/\hbar}$, which gives an exponent $f\sim \exp{-\im\bb{z^0(z_0- e\phi/c)+z^i(z_i- eA_i/c)}/\hbar}$, then
 \begin{align}
0&=\e\gamma\bb{1_\alpha^\gamma (\partial_{0}- \sfrac ec\phi)+\eta_\alpha^{\gamma\adj j}(\partial_j- \sfrac ecA_j)+\n H^\gamma_\alpha}f^\alpha\nonumber\\
&\propto\e\gamma\bb{1_\alpha^\gamma\sfrac1{\hbar} (-\im z_0- \sfrac ec\phi)+\eta_\alpha^{\gamma\adj j}\sfrac\im{\hbar}(-\im z_j- \sfrac ecA_j)+\n H^\gamma_\alpha}f^\alpha\nonumber\\
&\propto\e\gamma\bb{1_\alpha^\gamma (E-e\phi)+\eta_\alpha^{\gamma\adj j}\im(cp_j- eA_j)+\sfrac{\im\m c^2}{\hbar} H^\gamma_\alpha}f^\alpha
 \end{align}
giving the Dirac equation of a particle with charge $e$ in the presence of an electromagnetic four-potential $\phi+A$.

\subsection{The Maxwell equations}

The microscopic electromagnetic field equations can be viewed as simply the differentiability conditions \cref{anacond} in the natural geometry. 

Let $h=\phi^0\e0+A^i\e i$ be the electromagnetic 4-potential. Its derivative with respect to the usual spacetime 4-vector $z=t\e0+x^i \e i$ is
\begin{align}
f&=\bb{\sfrac{\partial\;}{\partial t}+\sfrac{\partial\;}{\partial{\bf x}}}\bb{\phi+{\bf A}}\nonumber\\
&=\bb{\partial_t\;\phi+\nabla\cdot{\bf A} }+\bb{\nabla\phi+\partial_t{\bf A} }+\nabla \times{\bf A}\nonumber\\
&=\hspace{1.1cm}\alpha\hspace{1.1cm}-\hspace{0.95cm}{\bf e}\hspace{0.9cm}+\hspace{0.4cm}\im\;{\bf B}\;,
\end{align}
defining an electric field $\bf E$, magnetic field $\bf B$, and some associated scalar field $\bf \alpha$. 

Let us consider the adjoint derivative of $f$ to vanish, that is $\sfrac{d\;}{d\adj z}f=0$, by \cref{anacond}. This gives
\begin{align}
0&=\bb{\sfrac{\partial\;}{\partial t}-\sfrac{\partial\;}{\partial{\bf x}}}(\alpha-{\bf E}+\im{\bf B})\nonumber\\
&=(\partial_t \alpha+\nabla \cdot{\bf E})\;+\;\im(\partial_t{\bf B}-\im\nabla \times{\bf E})\nonumber\\
&\quad\;\;\;\;-\;\im\nabla \cdot{\bf B}\;-\;(\partial_t{\bf E}+\im\nabla \times{\bf B}+\nabla \alpha)\;,
\end{align}
hence, splitting out the real and imaginary, scalar and vector, parts yields
\begin{equation}\label{max}
\begin{array}{rclrcl}
\nabla \cdot{\bf E}&\!=\!&-\partial_t \alpha\;,&\qquad
\im\nabla \times{\bf B}+\partial_t{\bf E}&\!=\!&-\nabla \alpha\;,\\
\nabla \cdot{\bf B}&\!=\!&0\;,&
\im\nabla \times{\bf E}-\partial_t{\bf B}&\!=\!&0\;.
\end{array}
\end{equation}
These obviously resemble the microscopic Maxwell equations, with the derivatives of $\alpha$ providing the charge and current sources terms, which appear only in the first row in correspondence with Maxwell's equations. 
(The factor of $\im$ ensures that the cross product is real-valued in this algebra, again consistent with the Maxwell equations). 

This does not suggest necessarily that $\alpha$ is actually an electromagnetic source term, i.e. the charge and current of the Maxwell equations, and the nature of $\alpha$ is unclear except that it arises in the electromagnetic fields as derived here. If we extract the $\gamma$ terms in the covariant derivatives of \cref{max}, the real parts of $\con(h)$ and imaginary parts of $\con(h)$ sit on the first and second rows of \cref{max}, respectively. 

Despite both charge and mass appearing in the previous section, we have seen so far nothing tying mass and energy together as source terms of curvature, but given the resemblance of these expressions to physical laws, it is tempting to speculate on what can be done using the key elements of the algebra, namely the mixture, and the inescapable role played by the imaginary $\im$. 


\subsection{Weak curvature solutions}\label{sec:weak}

In Einstein's theory, a particle in a gravitational field follows a geodesic (see e.g. \cite{edGRmath,E1916,schutz,wald}). This is a path where 
\begin{align}
u^\beta u^\alpha_{;\beta}&=u^\beta u^\alpha_{,\beta}+u^\beta u^\lambda\con^\alpha_{\lambda\beta}	
\end{align}
vanishes, with $u$ being the 4-momentum. 
In an electromagnetic field a particle no longer follows the geodesic according to the standard extensions to relativity, see e.g. \cite{wald}. 
In the above formalism we can extend this so that geodesic transport still applies in an electromagnetic field, without adding extra dimensions as in the Kaluza-Klein theory. 

Let us investigate this by perturbing the flat (Minkowski) metric. The Schwarzchild perturbation for a spherically symmetric gravitational field is
\begin{align}
g_{00}&=1/g^{00}=-1-\mu_g\psi\;,& g_{0j}&=g^{0j}=0\;,\\\nonumber 
g^{jj}&=\;\;g_{jj}\;\;=+1+\mu_g\psi\;,& g^{jk}&=g_{jk}=0,\quad j\neq k\;,
\end{align}
for some weak gravitional potential $\psi$, and a constant $\mu_g=2/c^2$ where $c$ is the speed of light. 

Before calculating the geodesic equation, let us try to include in this the effect of a weak electromagnetic field. We will do this by proposing that electromagnetism adds an imaginary perturbation to the time components of the metric, say as 
\begin{align}
g_{00}&=1/g^{00}=-1-\mu_g\psi-\im\mu_e\phi\;,& g_{0j}&=g^{0j}=\hf\im\mu A_j\;,\\\nonumber 
g^{jj}&=\;\;g_{jj}\;\;=+1+\mu_g\psi+\im\mu_e\phi\;,& g^{jk}&=g_{jk}=0,\quad j\neq k\;,
\end{align}
for some electromagnetic potential $\phi\ec0+A_j\ec j$, and some constant $\mu_e$ with units of $1/\sf Volts$. (These can easily be written directly as perturbations of the bases $\e\alpha$, but we will try to remain as close to standard theory as possible). 

Assume the connection is symmetric so that we can take the standard formula \cref{christoffel} (see also e.g. \cite{wald,misner73}), 
\begin{align}
\con^\sigma_{\alpha\mu}=\hf g^{\beta\sigma}(g_{\alpha\beta,\mu}+g_{\beta\mu,\alpha}-g_{\mu\alpha,\beta})\;.
\end{align}
We shall concentrate on the components with index $\mu=0$, which give
\begin{subequations}
\begin{align}
%
\con^i_{00}&= g^{\beta i}g_{0\beta,0}-\hf g^{\beta i}g_{00,\beta}\nonumber\\
\wk{&=-\hf g^{i i}g_{00,i}+g^{i i}g_{0i,0}
+\cc{\hf g^{0 i}g_{00,0}+g^{j i}g_{0j,0}-\hf g^{j i}g_{00,j}}_{j\neq i}\nonumber\\}
&= (\hf\mu_g\psi+\hf\im\mu_e\phi)_{,i}+\hf\im\mu_e A_{i,0}+\ord{A_i(\psi+\im\phi)_{,0}}\nonumber\\
&\approx \hf\mu_gG_i+\hf\im\mu_e E_i\\
\con^i_{j0}&=\hf g^{\beta i}(g_{j\beta,0}+g_{\beta0,j}-g_{0 j,\beta})\nonumber\\
\wk{&=g^{i i}g_{0[i,j]}
+\hf \cc{g^{i i}g_{ji,0}+ g^{j i}g_{jj,0}+ g^{0 i}g_{00,j}+ g^{k i}(g_{jk,0}+2g_{0[k,j]})}_{k\neq i\neq j}\nonumber\\}
&= \hf\im\mu_e A_{[i,j]}+\ord{A_i(\psi+\im\phi)_{,j}}\nonumber\\
&\approx \hf\im\eta^{ik}_j\mu_e B_k\\
\con^i_{i0}&=\hf g^{\beta i}(g_{i\beta,0}+g_{\beta0,i}-g_{0 i,\beta})\nonumber\\
\wk{&=g^{i i}g_{ii,0}+\hf \cc{ g^{0 i}g_{00,i}+ g^{k i}(g_{ik,0}+2g_{0[k,i]})}_{k\neq i,i}\nonumber\\}
&= -(\mu_g\psi+\im\mu_e\phi)_{,0}+\ord{A_i(\psi+\im\phi)_{,i}}
\end{align}
\end{subequations}
where $G=\nabla\psi$, $E=\nabla\phi+\partial_t A$, $B=\nabla\times A$, in the (perturbed) natural geometry. Note here the index $i$ is not summed over in the expression for $\con^i_{i0}$. 


The real and imaginary parts of these Christoffel symbols thus give the gravitational and electromagnetic fields, 
\begin{align}
\rep{\con^i_{00}}&=\hf\mu_gG_i  \;,&
\rep{\con^i_{(j0)}}&=0  \;,\nonumber\\
\imp{\con^i_{00}}&=\hf\mu_e E_i  \;,&
\imp{\con^i_{(j0)}}&=\hf\mu_e F^{ik}g_{kj}\;,
\end{align}
in terms of the dual $F^{\alpha\beta}$ of the electromagnetic tensor
\begin{align}\label{Fmaxwell}
F_{\alpha\beta}=\partial_{\alpha}A_\beta-\partial_{\beta}A_\alpha=A_{[\beta,\alpha]}\;.
\end{align}
Overall this $\con^\omega_{\gamma0}$ part of the connection is a Hermitian matrix
\begin{align}
\con^\omega_{\gamma0}&=(\;J^\omega_\gamma+\hf\im F^{\omega\lambda}\id_{\lambda\gamma}\;)\mu_e\;,
\end{align}
where
\begin{align}
J^\omega_\gamma=\mbox{\footnotesize$\bb{\begin{array}{cccc}0&G_1&G_2&G_3\\G_1&0&0&0\\G_2&0&0&0\\G_3&0&0&0\end{array}}  $}\;,\quad
F^{\omega\lambda}\id_{\lambda\gamma}=\mbox{\footnotesize$\bb{\begin{array}{cccc}0&E_1&E_1&E_3\\-E_1&0&-B_3&B_2\\-E_2&B_3&0&-B_1\\-E_3&-B_2&B_1&0\end{array}}  $}\;.
\end{align}


Before then working out the geodesic equation for a test particle of mass $m$ in this weak field, having introduced electromagnetism as an imaginary perturbation to the metric, we will consider our test particle to have a charge $e$. Suppose that this gives an imaginary perturbation to the energy component of the classical 4-momentum, as
\begin{align}
u^\alpha=\gamma\cc{(m+\im e\rho)c,m\v v}\;,
\end{align}
where $\gamma$ is the usual Lorentz factor, and $\rho$ is a small quantity with the units of $mass/charge$. 
For instance this constant might be $\rho=1/\sqrt{\eps_0{\sf G}}$ where $\eps_0$ is the vaccum permittivity and $\sf G$ is the gravitational constant, which would give $e\rho\sim6.6\times10^{-9}kg$ in SI units (with $e$ being the elementary unit of charge). 

Note that if this complex 4-momentum is the phase of a function $f=\exp{u^\alpha\e\alpha}$ (i.e. the imaginary part of the exponent of $f$ as suggested in \cref{ephi} and \cref{geodesiC}), then the phase $\im e\rho c\gamma$ is defined up to integer multiples of $2\pi$, hence $e=\sfrac{2\pi}{\rho c\gamma} n$ for $n=0,1,2,...$. 

Since we have both real and imaginary components we must ask how we define a geodesic, and we shall use the steepest descent result \cref{geodesiC}. Say a particle follows a path $x(\tau)$ with tangent vector $x'(\tau)=u$, and curvature $x''(\tau)=x'(\tau)\cdot\sfrac{d\;}{d x}x'(\tau)=u^\beta u_{;\beta}\e\alpha$. We define parallel transport as requiring only the vanishing of the real part of this, meaning there is no displacement of $u$ away from the path, while allowing the imaginary part to wander, permitting rotation of $u$ about the path (reminiscent of Weyl's early attempts to generalize Einstein's theory). We will therefore require the real part of
\begin{align}\label{geocom}
u^\beta u^\alpha_{;\beta}&=u^\beta u^\alpha_{,\beta}+u^\beta u^\lambda\con^\alpha_{\lambda\beta}\;,
\end{align}
to vanish. 

Expanding \cref{geocom}, and dividing by $m$, we have
\wf{\begin{align*}
u^0 u^i_{;0}+u^j u^i_{;j}&=
u^0 u^i_{,0}+u^j u^i_{,j}
+u^0 u^0\con^i_{00}	
+u^0 u^j\con^i_{j0}	
+u^j u^0\con^i_{0 j}
+u^k u^j\con^i_{j k}\nonumber\\
c(m+\im e\rho)m\dot v^i+m^2v^j v^i_{;j}&=
c(m+\im e\rho) mv^i_{,0}+m^2v^j v^i_{,j}
+c^2(m+\im e\rho)^2\con^i_{00}	\nonumber\\&\quad
+c(m+\im e\rho) mv^j\con^i_{j0}	
+mv^j c(m+\im e\rho)\con^i_{0 j}
+m^2v^k v^j\con^i_{j k}
\end{align*}\vspace{-0.6cm}}
\begin{align}
(m+\im e\rho)\dot v^i&=
(m+\im e\rho) \bb{ v^i_{,0}
+c(1+\im\sfrac{e\rho}m)\con^i_{00}	
+2v^j\con^i_{(j0)}}+\ord{v^2,v^i_{,j}}\;.
\end{align}
Vanishing of the real part gives
\begin{align}
-v^i_{,0}&=
c(1-\sfrac{e^2\rho^2}{m^2})\rep{\con^i_{00}}\;-\;2\sfrac {ec}m\rho\;\imp{\con^i_{00}}	\nonumber\\&\quad\;\;
+2v^j\rep{\con^i_{(j0)}}\;-\;2\sfrac em\rho v^j\imp{\con^i_{(j0)}}	\nonumber\\
&=\hf c\mu_g(1-\sfrac{e^2\rho^2}{m^2})\id^{ij}\psi_{,j} -\sfrac {e\rho}m\mu_e (\id^{ij}E_j + v^j \eta^{ik}_j B_k)	\nonumber\\
&=\hf c\mu_gG_i -\sfrac {ec\rho}m\mu_e (\id^{ij}E_j + v^j \eta^{ik}_j B_k)\;+\;\ord{\mu_g e^2\rho^2}\;.
\end{align}
In the natural geometry the magnetic term is the familiar cross-product $v^j \eta^{ik}_j B_k=(v\times B)_i$. 

Taking the gravitational component only we have the Newtonian force
\begin{align}
cmv^i_{,0}\approx-mG_i +\dots\;,\qquad\qquad\qquad\;\;
\end{align}
up to a higher order perturbation `$+\ord{e^2\rho^2}$' from the electromagnetic contribution to the particle's 4-momentum, with the `0' subscript denoting the derivative $v_{,0}=\sfrac1c\sfrac{\partial v}{\partial t}$.

Taking only the electromagnetic component we have
\begin{align*}
mv^i_{,0}\approx {ec\rho\mu_e} (E_i + v^j \eta^{ik}_j B_k)+\dots\;,
\end{align*}
which, if we let $\rho\mu_e=1/c^2$, provides the Lorentz force
\begin{align}
cmv^i_{,0}\approx {e} (E_i + v^j \eta^{ik}_j B_k)+\dots\;.\qquad
\end{align}
We note that this fixes the relation between $\rho$ or $\mu_e$, but leaves one of them undetermined.


\subsection{Electromagnetic field and stress-energy tensors}

To round off these physical relations we will remark on the simple and `natural' algebra of electromagnetic tensors in the mixture algebra. 
 
The electromagnetic fields and tensor can be related in various ways using the mixture, in particular
\begin{align}
\begin{array}{rl}
E^\gamma+\im B^\gamma=\eta^{\gamma\alpha}_\beta g^{\beta\delta} F_{\delta\alpha}
\end{array}
\end{align}
or $\v E+\im\v B=\e\gamma\eta^{\gamma\alpha}_\beta g^{\beta\delta} F_{\delta\alpha}$, 
and
\begin{align}
\eta^{\mirror\alpha\beta}_\gamma(\v E+\im\v B)^\gamma&=- F_{\alpha\beta}- \im G_{\alpha\beta}
\end{align}
where $G_{\alpha\beta}$ is the dual Faraday tensor. 

The relation between the electromagnetic field, the stress-energy tensor, and the Poynting vector are particularly simple. Taking the product of the complex electromagnetic field with its conjugate, we have
\begin{align}
(\v E+\im\v B)(\v E-\im\v B)&=(\v E+\im\v B)^\alpha(\v E-\im\v B)^\beta\eta_{\alpha\beta}^\gamma\e\gamma\nonumber\\
&= T^{\alpha\beta}\eta_{\alpha\beta}^\gamma\e\gamma\nonumber\\
&= Q^\gamma\e\gamma\;=\; Q\;,
\end{align}
where $Q$ is the Poynting vector, and $T$ is the usual electromagnetic stress-energy tensor $T^{\alpha\beta}=F^{\alpha\mu}F^\beta_\mu-\sfrac14g^{\alpha\beta}F_{\mu\nu}F^{\mu\nu}$. 

Just for completeness, we include that the similar quantity for the stress-energy tensor of a perfect fluid in thermodynamic equilibrium, $T^{\alpha\beta}=(\rho+p)u^\alpha u^\beta+pg^{\alpha\beta}$
, gives 
\begin{align}
 T^{\alpha\beta}\eta_{\alpha\beta}^\gamma\e\gamma&=(\rho+p)u^2+2p\e0\;.
\end{align}

\section{Closing remarks, and a complex variable for physical geometry}\label{sec:close}

We have arrived at a schema of algebra and calculus where all quantities, particularly products, ratios, and derivatives, belong to the same algebraic space. Founding covariant calculus upon the `mixture' of bases permits us to extend analyticity to geometric algebras beyond the complex variables. It even allows us to consider non-associative algebras, with the loss of the index exchange lemma \cref{etaexchange}, with much of the theory above still applying. 

The geometry of our physical world is obviously associated in some manner with the algebra of the quaternions. However, as Hamilton and many since have encountered, as we go from basic algebra into functions and then to calculus, their aesthetic simplicity begins to degrade and succumbs eventually to insurmountable obstructions. Although geometric algebras circumvent many of those obstructions, in doing so they give up some of the simpler character of Hamilton's original formulations. The {\it mixture} re-instills this character, and at the same time lends itself naturally to the powerful rigours of tensor calculus. 


Although I have used fundamental physics to explore and illustrate this formalism, I have not attempted to lay out a physical theory, merely to use the idea of mixtures and analytic curvature to ask how much of physical law might stem from our natural instinct to find empirical quantities with simple mathematical properties --- even before we understand the nature of those properties. Our initial problem of seeking to extend certain integration methods to higher dimensions led us to ask, 
if one seeks to describe empirical phenomena by quantities that satisfy simple differential relations, how much of the character of physical law is then mathematically inevitable? 
%
The framework set out here no doubt has flaws and holes, and purely from the mathematical side, in seeking an alternate path towards differential geometry ignores parallel developments in the subject over the last century. I have presented it as such to avoid unecessary abstractions, and avoid the pretence of rigour. 
Insofar as we have explored this algebra, we have barely touched upon the symmetries and group properties it permits in physical relations, and this may be one of the key areas for future investigation. 


Since we have played somewhat lightly with physical laws here, let me conclude with a conjecture to be considered even more lightly. 
The quaternion algebra invites us to combine the quantities of space and time into a single space-time four-vector, as well as energy and momentum into a single energy-momentum four-vector. The use of imaginaries further invites us to combine these into a single complex four-vector of space-time-energy-momentum. 
Assume a dimensionless complex quaternion 
$$q=a_0t+a_1{\bf x}+\im(a_2E+a_3{\bf p})$$
where $\tau$, $\bf x$, $E$, and $\bf p$ represent time, displacement, energy, and momentum respectively. The coefficients $a_i$ provide the necessary dimensional units to make such an expresison balance. If we assume these are functions of the fundamental constants of the speed of light $c$, Planck's constant $\hbar$, and gravitational constant $G$, by dimensional arguments we arrive at $$\cc{a_0,a_1,a_2,a_3}=\cc{c,1,\sfrac{L^2}{\hbar c^2},\sfrac{L^2}{\hbar c}}/L$$
in terms of the Planck length $L=\sqrt{\hbar G/c^3}$, so $q$ is otherwise written
$$q=\sfrac1L\bb{ct+{\bf x}}+\sfrac{\im L}{\hbar c}\bb{E/c+{\bf p}}\;.$$
If we consider the length
$$|q|^2L^2=\bb{c^2t^2-{\bf x}^2}-\sfrac{L^4}{\hbar^2c^2}(E^2/c^2-{\bf p}^2)+2\sfrac{\im L^2}{\hbar c}\bb{tE+{\bf x}\cdot{\bf p}}\;,$$
then factors of different powers in the small parameter ${L^2}/{\hbar c}\sim10^{-45}/k\hspace{-0.04cm}gms^{-2}$ decouple into three asymptotically (as ${L^2}/{\hbar c}\rightarrow0$) independent quantities
$$c^2t^2-{\bf x}^2,\qquad tE+{\bf x}\cdot{\bf p},\qquad E^2/c^2-{\bf p}^2,\;$$
namely the spacetime geodesic interval (squared), the optical path length, and the rest mass (squared). Being of such different orders of magnitude we may expect these to behave quasi-independently on relative timescales $1:{L^2}/{\hbar c}:{L^4}/{\hbar^2c^2}$. 
%
This suggests the possibility of devising a dynamical system in which the relative division of these regimes, into a patchwork of separate spacetime, inertial, and electromagnetic phenomena, can be better understood. But in this, we have certianly strayed beyond the remit of this paper. 

To return to the mathematics, we might ask to what extent the familiar symbolic rules of differentiation can be carried into higher dimensions using the mixture and analytic curvature, for instance whether we can find a space in which, for a constant $p$, we have $\sfrac{d\;}{dz}pz=p$, $\sfrac{d\;}{dz}z^p=pz^{p-1}$, $\sfrac{d\;}{dz}e^{pz}=e^{pz}$, and so on. To some extent we can, but as far as the author has found, we can do so only locally. A connection can be found such that the general series 
$$\sum_{p=0}^\infty a_p(z-z_0)^p$$
is differentiable, integrable, and satisfies the familiar symbolic rules of differentiation, but doing so such that the function and all its derivatives are scalars at $z=z_0$. This seems rather limiting and so we omit it here, but it may deserve further attention. 

Our deeper study should from here proceed into the geometry of functions using mixture algebra, the topology of stationary phase and steepest descent curves, and resulting methods of integration, in which the toughest challenge is in finding the connection term $\con$ for general functions and its integrals in spaces `curved by analyticity'. These we leave to future work.

\appendix

\section{Appendix}
\subsection{Deriving the natural geometry}\label{sec:Anatgeo}

As an example algebra let us take the geometry of everyday mechanics, namely that of displacements and rotations. We deviate in two key respects from the usual description of quaternions or Clifford algebras. 

The first deviation is how we define the product of two basis vectors $\e\alpha$ and $\e\beta$, via a {\it mixture} tensor $\eta$, such that $\e\alpha\e\beta=\eta_{\alpha\beta}^\gamma\e\gamma$. 
The second deviation is how this mixture will distinguish displacements from rotations, namely if we denote a displacement $x$ along some axis $\e\alpha$ as $x\e\alpha$, then we denote a rotation $\omega$ about that axis as $\im\omega\e\alpha$ in terms of the imaginary $\im=\sqrt{-1}$. 

The arguments behind the algebraic relations in \cref{sec:natgeo} are only a slight variation of those of W. R. Hamilton in \cite{H1846}. 
They are based on taking, as a starting point, a planar geometry described by {\it real} basis vectors $\e1$ and $\e2$, of unit length. 
Being real, these are taken to be their own inverses, $\e i=1/\e i$. 
(This is contrary to Hamilton's choice for the quaternion vector bases which would satisfy $\e i=-1/\e i$, which we will consider `imaginary' and will derive shortly, but otherwise the argument hereon differs little from Hamilton's.) 

Let the ratio of two lengths along the same direction $\e i$ be measured along a scalar basis $\e0$. As a real scalar, this basis is also its own inverse, $1/\e0=\e0$. Moreover since it is a unit scalar it behaves as the identity, therefore $\e i/\e0=\e i$ for $i=1,2$. 

The ratio of lengths along orthogonal directions $\e1$ and $\e2$ cannot then lie along either $\e0$, $\e1$, or $\e2$, otherwise $\e1$ and $\e2$ are not orthogonal. Let the operation relating $\e1$ to $\e2$ therefore be measured along a new basis $\omegab=\e1/\e2$, a `rotation' basis. The rotation from $\e1$ to $\e2$ is the opposite to that from $\e2$ to $\e1$, such that $\e1/\e2=-\e2/\e1$, therefore the inverse of $\omegab$ is $1/\omegab=-\omegab$. This introduces our first imaginary quantity, and one of Hamilton's quaternion bases (or bivector in modern terminology). We can readily show that $\e1,\e2,\omegab,$ anticommute with each other, but commute with $\e0$. Therefore let us call this new direction $\omegab=\im\e3$, i.e. lying along a new real direction $\e3$ but with an imaginary factor so that $1/\omegab=1/(\im\e3)=-\im\e3$. 

Thus the bases $\e1,\e2,\e3,$ are identified with distinct directions in space. The system on $\e0,\e1,\e2,\e3$, is closed under multiplication, division, addition, and subtraction, and provides a description of displacements and, using the imaginary, of rotations. 
The composition of a $\sfrac\pi2$-rotation about the $1$-axis and the $2$-axis is a $\sfrac\pi2$-rotation about the $3$-axis, as is verified by calculating $\ec1\ec2=\e2\e3\e3\e1=\e2\e1=-\ec3$, and similarly $\ec2\ec3=-\ec1$, $\ec3\ec1=-\ec2$. 

The unit $\im$ commutes with all of the bases $\e\alpha$ and $\ec\alpha$ for $\alpha=0,1,2,3$, so must be a scalar itself. Furthermore note that $\e1\e2\e3=\e1\ec1=\im$. 

Thus we arrive at the relations 
$(\e\alpha)^2=-(\im\e\alpha)^2=\e0\;,$ $\e1\e2\e3=\im\;,$ and $\e i\e j=-\e j\e i=\im\e k\;,$ (where $\cc{i,j,k}$ are a cyclic permutation of $\cc{1,2,3}$), 
and hence to the complex quaternions as a natural algebra for geometry. The quaternions correspond to taking bases $\cc{\e0,\im\e1,\im\e2,\im\e3}$, replacing what are sometimes called bivectors or rotors with imaginary vectors $\im\e k$. The natural geometry can instead be described as consisting of a rectilinear set of orthogonal bases $\cc{\e\alpha}_{\alpha=0,1,2,3}$, such that real scalars are measured along $\e0$, and real displacements are measured along $\cc{\e i}_{i=1,2,3}$, while imaginaries denote rotations about those directions, and so we obtain \cref{hamilton}.  

\bigskip
\subsection{$\eta$ identities}\label{sec:Amixture}

Certain useful symmetries and other properties of the mixture tensor follow from its definition. 

Firstly is the cyclic permutation of indices:
\begin{align}
\eta_{\alpha\beta}^\gamma=\eta_{\beta\gamma}^\alpha=\eta_{\gamma\alpha}^\beta\quad\mbox{and}\quad\eta^{\alpha\beta}_\gamma=\eta^{\beta\gamma}_\alpha=\eta^{\gamma\alpha}_\beta\;.
\end{align}
The complex conjugate, by \cref{def:conj}, acts on the mixture as
\begin{align}
\eta_{\alpha\beta}^\gamma=\conj{(\eta_{\beta\alpha}^\gamma)}\quad\mbox{and}\quad\eta^{\alpha\beta}_\gamma=-\conj{(\eta^{\beta\alpha}_\gamma)}\;.
\end{align}
The mixture $\eta_{\alpha\beta}^\gamma$ has a pseudo-inverse, $\eta^{\beta\alpha}_\delta$, 
\begin{align}
\eta_{\alpha\beta}^\gamma\eta^{\beta\alpha}_\delta=\id_\delta^\gamma\quad\mbox{and}\quad
\eta_{\alpha\beta}^\gamma\eta^{\delta\alpha}_\gamma=\id^\delta_\beta\;.
\end{align}
If the algebra is associative, the mixture satisfies certain index-exchange identities, 
\begin{align}\label{etaexchange}
\begin{array}{rl}
&\eta_{\gamma\lambda}^\beta\eta^{\alpha\lambda}_\delta=\eta^{\alpha\beta}_\lambda\eta_{\gamma\delta}^\lambda\qquad\&\qquad
\eta_{\gamma\lambda}^\beta\eta^{\lambda\alpha}_\delta=\eta^{\alpha\lambda}_\gamma\eta_{\lambda\delta}^\beta\\
&\eta_{\lambda\gamma}^\beta\eta^{\lambda\alpha}_\delta=\eta^{\beta\alpha}_\lambda\eta_{\delta\gamma}^\lambda\qquad\&\qquad
\eta_{\lambda\gamma}^\beta\eta^{\alpha\lambda}_\delta=\eta^{\lambda\alpha}_\gamma\eta_{\delta\lambda}^\beta
\end{array}
\end{align}
and triple-indentities, 
\begin{align}
\begin{array}{rl}
&\eta_{\beta\gamma}^\alpha=\sfrac1n\eta_{\mu\nu}^\alpha\eta_{\beta\omega}^\mu\eta^{\nu\omega}_\gamma\qquad\&\qquad\eta^{\beta\gamma}_\alpha=\sfrac1n\eta^{\mu\nu}_\alpha\eta^{\beta\omega}_\mu\eta_{\nu\omega}^\gamma\\
&\eta_{\gamma\beta}^\alpha=\sfrac1n\eta_{\nu\mu}^\alpha\eta_{\omega\beta}^\mu\eta^{\omega\nu}_\gamma\qquad\&\qquad\eta^{\gamma\beta}_\alpha=\sfrac1n\eta^{\nu\mu}_\alpha\eta^{\omega\beta}_\mu\eta_{\omega\nu}^\gamma
\end{array}
\end{align}
in $n$ dimensions. 

The properties above are quite straightforward to show. The proof of the index exchanges follows as
\begin{align}
\eta_{\gamma\lambda}^\beta\eta^{\alpha\lambda}_\delta\ec\delta
&=\eta_{\gamma\lambda}^\beta\ec\alpha\ec\lambda
=\ec\alpha\ec\beta\e\gamma\nonumber\\
&=\eta^{\alpha\beta}_\lambda\ec\lambda\e\gamma
=\eta^{\alpha\beta}_\lambda\eta_{\gamma\delta}^\lambda\ec\delta
\quad\mbox{etc.}
\end{align}
The triple-identities can be shown using a combination of index-exchange and the inverse, or directly using the basis,  
\begin{align}
\eta_{\mu\nu}^\alpha\eta_{\beta\omega}^\mu\eta^{\nu\omega}_\gamma\e\alpha
&=\e\mu\e\nu\eta_{\beta\omega}^\mu\eta^{\nu\omega}_\gamma
=(\e\beta\e\omega)(\ec\omega\e\gamma)\nonumber\\
&=n\e\beta\e\gamma
=n\eta_{\beta\gamma}^\alpha\e\alpha\quad\mbox{etc.}
\end{align}
As well as \cref{etainverse}, various other pseudo-inverses can be found, such as $\sfrac{1}s\eta_{\beta\gamma}^\alpha\eta^{\mirror\gamma\beta}_\delta=1^\alpha_\delta$ where $s$ is the signature of the basis, $s=\e\alpha\ec{\mirror\alpha}$. 

\bigskip

\bibliography{../grazcat}

\begin{thebibliography}{10}

\bibitem{cartan1922}
E.~Cartan.
\newblock Sur une g\'en\'eralisation de la notion de courbure de riemann et les
  espaces \`a torsion.
\newblock {\em C. R. Acad. Sci.}, 174:593--595, 1922.

\bibitem{cg07}
W.~N. Cottingham and D.~A. Greenwood.
\newblock {\em An introduction to the standard model of particle physics}.
\newblock Cambridge University Press, 2007.

\bibitem{dirac28}
P.~A.~M. Dirac.
\newblock The quantum theory of the electron.
\newblock {\em Proc. R. Soc. A}, 1928.

\bibitem{diracqm}
P.~A.~M. Dirac.
\newblock {\em The Principles of Quantum Mechanics}.
\newblock Oxford University Press, 1930.

\bibitem{dirac1931}
P.~A.~M. Dirac.
\newblock Quantised singularities in the electromagnetic field.
\newblock {\em Proc. R. Soc. A}, 133:60--72, 1931.

\bibitem{geoma}
C.~Doran and A.~Lasenby.
\newblock {\em Geometric Algebra for Physicists}.
\newblock Cambridge University Press, 2007.

\bibitem{edGRmath}
A.~S. Eddington.
\newblock {\em The mathematical theory of relativity}.
\newblock Cambridge University Press, 1923.

\bibitem{E1916}
A.~Einstein.
\newblock {Die Grundlage der allgemeinen Relativit\"atstheorie}.
\newblock {\em Annalen der Physik}, 49(7):769--822, 1916.

\bibitem{H1846}
W.~R. Hamilton.
\newblock On symbolic geometry.
\newblock {\em Cambridge and Dublin Mathematical Journal}, i/ii/iii/iv:in vol.
  i: p45--57,137--154,256--263, in vol. ii: p47--52,130--133,204--209, in vol.
  iii:p68--84,220--225, in vol. iv:p84--89,105--118, 1846/7/8/9.

\bibitem{levicivita}
T.~Levi-Civita.
\newblock {\em The Absolute Differential Calculus}.
\newblock Dover (Transl. 1927 by Miss M Young of Lezioni di calcolo
  differenziale assoluto 1925), 2013.

\bibitem{misner73}
C.~W. Misner, K.~S. Thorne, and J.~A. Wheeler.
\newblock {\em Gravitation}.
\newblock San Francisco: W. H. Freeman, 1973.

\bibitem{dawn}
L.~O'Raifeartaigh.
\newblock {\em The Dawning of Gauge Theory}.
\newblock Princeton University Press, 1997.

\bibitem{schutz}
B.~F. Schutz.
\newblock {\em A first course in general relativity}.
\newblock Cambridge University Press, 1985.

\bibitem{wald}
R.~M. Wald.
\newblock {\em General Relativity}.
\newblock University of Chicago Press, 1984.

\bibitem{yangmills}
C.~N. Yang and R.~K. Mills.
\newblock Conservation of isotopic spin and isotopic gauge invariance~.
\newblock {\em Physical Review}, 96(1):191--5, 1954.

\end{thebibliography}
\bibliographystyle{plain}

\end{document}